\documentclass[iop]{emulateapj}
\usepackage{epstopdf}
\usepackage{natbib}
\usepackage{textcomp}
\usepackage{hyperref}
\usepackage{graphicx,color}
\usepackage{times}
\usepackage{amssymb}
\usepackage{url}
\usepackage{amsmath}
\usepackage[english]{babel}

\newcommand{\sdsszeroeight}{SDSS J082625.70+612515.10}
\newcommand{\sdssonethree}{SDSS J134144.60+474128.90}

\newcommand{\sdsszero}{SDSS~J0826+6125}
\newcommand{\sdssone}{SDSS~J1341+4741}

\shorttitle{Two EMP stars from SDSS-MARVELS pre-survey} 
\shortauthors{Bandyopadhyay et al.}

\begin{document}

\title{Chemical composition of two bright extremely metal-poor stars from the\\
SDSS MARVELS pre-survey}

\author{Avrajit Bandyopadhyay\altaffilmark{1}, Sivarani Thirupathi\altaffilmark{1}, Antony Susmitha\altaffilmark{1},
 Timothy C. Beers\altaffilmark{2},  Sunetra Giridhar\altaffilmark{1}, Arun Surya\altaffilmark{1}, Thomas Masseron\altaffilmark{3}}

% \author{Susmitha Rani Antony\altaffilmark{1}}

\email{avrajit@iiap.res.in,sivarani@iiap.res.in}

\altaffiltext{1}{Indian Institute of Astrophysics, Bangalore}
\altaffiltext{2}{Department of Physics and JINA Center for the Evolution
of the Elements, University of Notre Dame, Notre Dame, IN, 46656, USA}
\altaffiltext{3}{Departmento de Astrofisica, Universidad de La Laguna, E-38206 La Laguna, Tenerife, Spain}

\begin{abstract}
\sdsszeroeight\ (V = 11.4; [Fe/H] = $-$3.1) and
\sdssonethree\ (V = 12.4; [Fe/H] = $-$3.2) were observed with the SDSS
2.5-m telescope as part of the SDSS-MARVELS spectroscopic pre-survey,
and were identified as extremely metal-poor (EMP; [Fe/H] $< -3.0$) stars
during high-resolution follow-up with the Hanle Echelle Spectrograph
(HESP) on the 2.3-m Himalayan Chandra Telescope. In this paper, the
first science results using HESP, we present a detailed analysis of
their chemical abundances. Both stars exhibit under-abundances in their
neutron-capture elements, while one of them,
\sdssonethree, is clearly enhanced in carbon. Lithium was also detected in
this star at a level of about A(Li) = 1.95. The spectra were
obtained over a span of 6-24 months, and indicate that both stars could
be members of binary systems. We compare the elemental abundances
derived for these two stars along with other carbon-enhanced
metal-poor (CEMP) and EMP stars, in order to understand the nature of
their parent supernovae. We find that CEMP-no stars and
EMP dwarfs exhibit very similar trends in their lithium abundances at various
metallicities. We also find indications that CEMP-no stars have larger
abundances of Cr and Co at a given metallicity, compared to EMP
stars.
\end{abstract}


\section{Introduction}
Extremely metal-poor (EMP; [Fe/H] $< -3.0$) stars of the Galactic halo
are thought to be the immediate successors of the first stars, and were
likely to have formed when the Universe was only a few hundred million
years old (e.g., \citealt{bromm2009}) -- their evolution and
explosion led to the first production of heavy elements.
These first supernovae had a considerable dynamical, thermal, and
chemical impact on the evolution of the surrounding interstellar medium,
including mini-halos that can be some distance away from the location of
the first-star explosion \citep{cooke2014,smith2015,chiaki2018}. Stars
(and their host galaxies) that formed thereafter are expected to carry
the imprints of the nucleosynthesis events from these Population III
stars \citep{beers2005,frebelandnorris,sharma2018}. Studies of such EMP
stars have greatly benefited from the large spectroscopic surveys that
were carried out in the past in order to identify them in significant
numbers, such as the HK survey of Beers and collaborators
\citep{beers85,beers92} and the Hamburg/ESO Survey of Christlieb and
colleagues \citep{christlieb2003}. More recent surveys, such as SDSS,
RAVE, APOGEE and LAMOST continue to expand the known members of this
rare class of stars (e.g., \citealt{rave2010fulbright};
\citealt{ivezic2012}; \citealt{aokibeers2013}; \citealt{zhao2012lamost}; 
\citealt{apogee2013}; \citealt{apogee2014}; \citealt{li2015lamost}). 

High-resolution spectroscopic studies of metal-poor Galactic halo stars
have demonstrated diversity in their chemical compositions. For example,
on the order of 20\% of stars with [Fe/H] $< -2.0$ exhibit large
enhancements in their carbon-to-iron ratios ([C/Fe] $> +0.7$;
\citealt{aokibeers2007}; \citealt{lee2013}; \citealt{lee2017}). As shown 
by numerous studies, the frequency of carbon-enhanced metal-poor (CEMP)
stars continues to increase with decreasing [Fe/H]. The fractions of
CEMP stars also increase with distance from the Galactic plane
\citep{frebel2006,beers2017}, and also between the inner-halo and
the outer-halo regions \citep{lee2017}.  

CEMP stars can be separated into four sub-classes \citep{beers2005}: i)
CEMP-$s$ stars, which show enhancements of $s$-process elements, ii)
CEMP-$r$ stars, which exhibit enhancements of $r$-process elements, iii)
CEMP-$r/s$ stars, which show enhancements in both $r$- and $s$-process
elements\footnote{\cite{hampel2016} suggest that the observed heavy
element patterns of these stars are well accounted for by an
``intermediate neutron-capture process,'' (as first suggested by
\citealt{cowanandrose}), and should be referred to henceforth as CEMP-$i$
stars.}, and iv) CEMP-no stars, which exhibit no neutron-capture element
enhancements. Long-term radial-velocity monitoring studies have shown
that most ($> 80$\%, possibly all) CEMP-$s$ stars are members of binary
systems involving a (now extinct) asymptotic giant branch (AGB) star
that transferred carbon and $s$-process rich material to the presently
observed (lower-mass) star \citep{lucatello,starkenburg2014,
hansen2016a}, while CEMP-no stars exhibit observed binary frequencies
typical of non-carbon-enhanced halo giants, $\sim 18$\%
\citep{starkenburg2014,hansen2016b}.

\cite{jinmiyoon} have considered the rich morphology of the absolute
abundance of carbon, $A$(C)$ = log(C)$, as a function of [Fe/H], based on
high-resolution analyses of a large sample of CEMP stars (their Figure
1, the Yoon-Beers diagram). In addition to their Group~I stars, which
are dominated by CEMP-$s$ stars, they demonstrate that the CEMP-no stars
not only exhibit substantially lower $A$(C), but bifurcate into two
apparently different regions of the diagram, which they refer to as
Group~II and Group~III stars. This behavior immediately suggests that
these groups might be associated with different progenitors responsible
for the carbon production, a suggestion borne out by the modeling
carried out by \cite{placco2016}, and/or on the masses of the mini-halos
in which these stars formed. \citet{chiaki2017} have emphasized that
different cooling pathways, dependent on the formation of carbon or
silicate dust, may have applied to the Group~III and Group~II stars in
the Yoon-Beers diagram.

Multiple models for the production of CEMP-no stars have been considered
in the literature, such as the ``spinstar'' models (e.g.,
\citealt{meynet2006}; \citealt{meynet2010}; \citealt{chiappini2013}), and the ``mixing and
fallback'' models for faint SNe (e.g., \citealt{umedanomoto2003}; 
\citealt{umedanomoto2005}; \citealt{nomoto2013};  
\citealt{tominaga2014}). Both processes may well play a role
\citep{maeder2015,choplin}.

Regardless of the complexity of the situation, additional detailed
observations of EMP stars with and without clear carbon enhancement,
such as those carried out here, are required for progress in
understanding.  This paper is outlined as follows.  In Section 2 we
describe our high-resolution observations.  Consideration of possible
radial-velocity variations for our two targets is presented in Section 3.
Section 4 summarizes our estimates of stellar atmospheric parameters,
and describes our abundance analyses.  Results of the abundance analysis
are reported in Section 5.  We present a discussion of our results with a comparative study of 
CEMP-no and EMP stars in
Section 6, along with a brief conclusion in Section 7.

\section{Observations and Analysis}
\subsection{Sample Selection}
MARVELS \citep{paegertmarvels}, a multi-object radial-velocity survey
designed for efficient exo-planet searches, was one of the three
sub-surveys carried out as part of SDSS-III \citep{eisenstein}. The
targets for the first two years of MARVELS were selected based on a
lower-resolution ($R \sim 1800)$ spectroscopic pre-survey using the SDSS
spectrographs. Most of the pre-survey observations were carried out
during twilight, when the fields were at low elevation. Targets were
selected from these pre-survey fields for the MARVELS main radial velocity (RV) survey,
which were later observed at higher elevations. There were about 30,000
stars observed as part of the spectroscopic pre-survey of stars with
$B-V > 0.6$ and $8 < V < 13$. Target fields for the first two years of
the MARVELS survey were around known RV standards, and about 75\% of the
target fields were in the Galactic latitude range $2^{\circ} < |b| <
30^{\circ}$. Although not the ideal location to find metal-poor stars,
it does offer the chance to identify a small number of bright halo
targets, suitable for high-resolution spectroscopic follow-up with
moderate-aperture telescopes, that happen to fall into the MARVELS
pre-survey footprint during their orbits about the Galactic center. The
pre-survey also has simple magnitude and color cuts, which reduces potential
selection biases. As in our previously published work
\citep{susmitha}, we used synthetic spectral fitting of the pre-survey
data to identify new metal-poor candidates. Here, we present
high-resolution observations and analysis of two EMP stars,
\sdsszeroeight\ (hereafter, \sdsszero) and \sdssonethree\ (hereafter,
\sdssone), with $V$ magnitudes of 11.44 and 12.38, respectively. These
two stars were selected for follow up, as they were found to be very
metal poor from spectral fitting of the pre-survey data, and were
also very bright. Results from the spectral fitting used to identify
metal-poor candidates from the MARVELS pre-survey will be discussed in a
separate paper.

\subsection{High-Resolution Observations}
High-resolution ($R \sim 30,000$) spectroscopic observations of the two EMP
stars were obtained with the Hanle Echelle Spectrograph (HESP) on the
2.3-m Himalayan Chandra telescope (HCT) at the Indian Astronomical
Observatory (IAO). The dates of observation, wavelength coverage, radial
velocities, and signal-to-noise ratios of the available spectra are
listed in Tables 1 and 2.

Data reduction was carried out using the IRAF\footnote{IRAF is
distributed by the National Optical Astronomy Observatory, which is
operated by the Association of Universities for Research in Astronomy
(AURA) under cooperative agreement with the National Science
Foundation.} echelle package. HESP has a dual fiber mode available, one
fiber for the target star and another, which can be fed with a
calibration source for precise RV measurements, or by the
night sky through a pinhole that has a separation of about 13\arcsec\
from the target. The sky fiber is used for background subtraction. All
the orders were normalized, corrected for radial velocity, and merged to
produce the final spectrum. The equivalent widths for individual species
are listed in the tables in the appendix. Recently, a custom data
reduction pipeline has been developed by A. Surya (available
publicly\footnote{$ https://www.iiap.res.in/hesp/ $})
that is more suitable for the crowded and curved orders of the stellar
spectra observed with HESP. However, in the present paper, we use IRAF,
and proper care was taken to avoid drift of the spectral tracings
blending into adjacent orders.

\begin{table}
\begin{center}
\caption{Observation log and radial velocities for SDSS~J0826+6125} \label{tbl-1}
\begin{tabular}{ccccrrrrrrr}
\tableline\tableline
Date & MJD & $\lambda$ Coverage & SNR & Radial Velocity \\
     &     & (\AA)              &     & (km~sec$^{-1}$) \\
\tableline
2015-11-03 &57330.20903 &3600-10800 &51 &$-$110.4 \\
2015-11-29 &57356.36042 &3600-10800 &49 &$-$95.6 \\
2015-12-22 &57379.10417 &3600-10800 &47 &$-$80.3 \\
2016-01-27 &57415.09792 &3600-10800 &47 &$-$52.3 \\
2016-10-20 &57682.21667 &3600-10800 &50 &$-$108.9 \\
2016-11-16 &57709.13542 &3600-10800 &51 &$-$104.1 \\
\tableline
\end{tabular}
\end{center}
\end{table}

\begin{table}
\begin{center}
\caption{Observation log and radial velocities for SDSS~J1341+4741} \label{tbl-2}
\begin{tabular}{cccccrrrrrrr}
\tableline\tableline
Date & MJD & $\lambda$ Coverage & SNR & Radial Velocity \\
     &     & (\AA)              &     & (km~sec$^{-1}$) \\
\tableline
2016-27-01 &57415.24722 & 3600-10800 &43 &$-$240.1\\
2016-24-04 &57503.18819 & 3600-1080 &49 &$-$190.5\\
2016-26-04 &57505.06458 & 3600-1080 &47 &$-$192.1\\
2016-24-06 &57564.02361 & 3600-1080 &48 &$-$176.2\\
2016-25-06 &57565.20139 & 3600-1080 &47 &$-$174.5\\
\tableline
\end{tabular}
\end{center}
\end{table}

\section{RADIAL VELOCITIES}
The HESP instrument is thermally controlled to ${\Delta}T$ = $\pm$ 0.1$^\circ$~C
at a set point of 16$^\circ$~C over the entire year, which is expected to
provide a long-term stability of $\sim 200$ m~s$^{-1}$ (Sivarani et al.
in preparation), substantially lowering systematic errors with respect
to a spectrograph that does not have such control. 

RVs were calculated for \sdsszero\ based on six epochs
of observations spread over a period of 12 months. For \sdssone, we
obtained five observations spread over six months. A cross-correlation
analysis was performed with a synthetic template spectrum suitable for
each star to obtain the RV measurement for each spectrum. We made use of
the software package RVLIN provided by \citet{rvlin}, which is a set of
IDL routines used to fit Keplerian orbits to derive the orbital
parameters from the RV data. The RV measurements exhibit peak-to-peak
variations of $\sim 60$ km~s$^{-1}$, with a period of 180 days for
\sdsszero, and $\sim 110$ km~s$^{-1}$, with a period of 116 days for
\sdssone. The best-fit orbits for these stars, based on the data in hand,
are shown in Figures 1 and 2. Although the existence of RV variations is
secure, with such sparse coverage of the proposed orbits more data are
required to confirm the periods.

\begin{figure}[!tbp]
\centering
\begin{minipage}[b]{0.4\textwidth}
    \includegraphics[width=\textwidth]{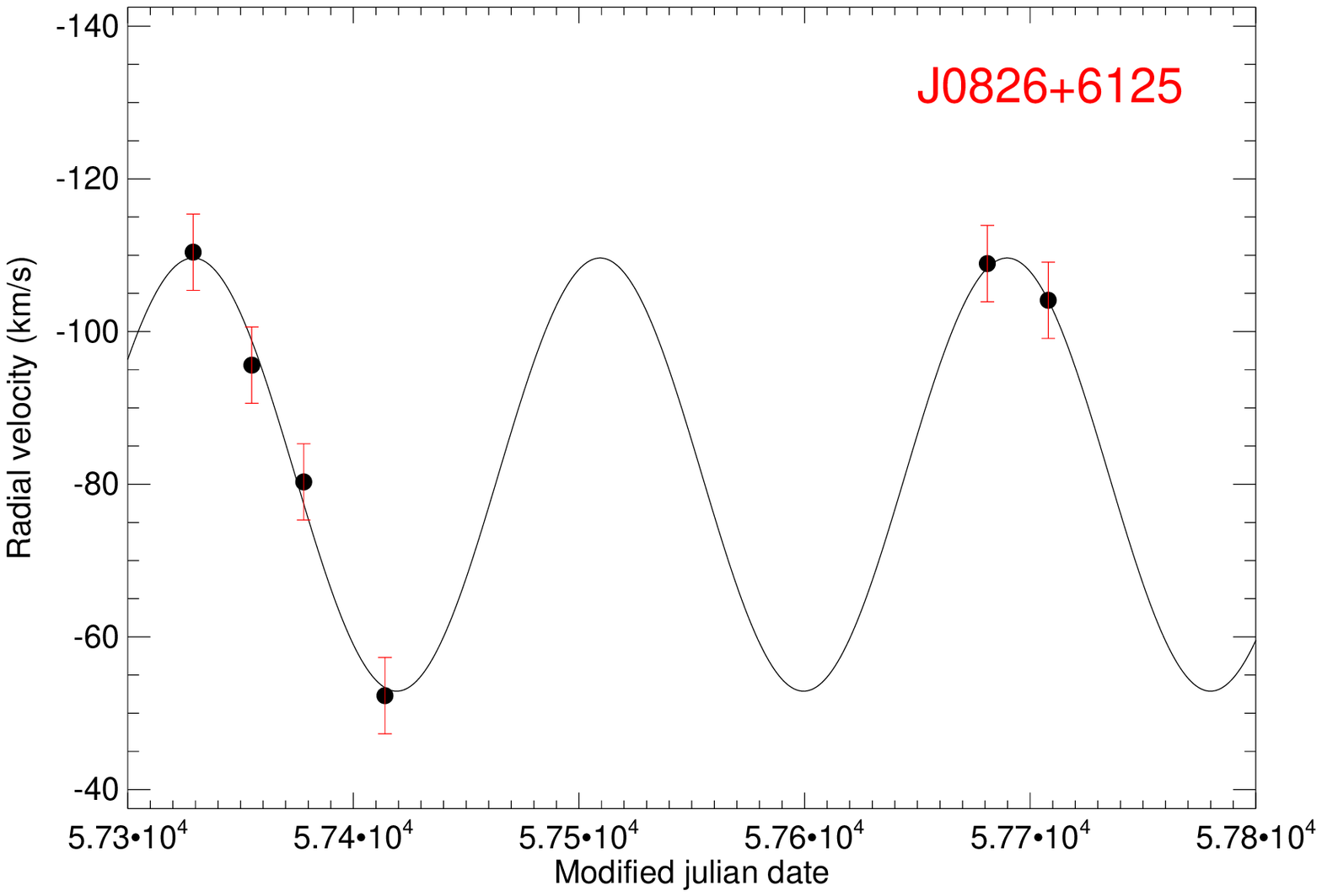}
    \caption{\normalfont \small Variation of radial velocity for \sdsszero. The derived period is
of 180.4 days} 
 \end{minipage}
  \hfill
  \begin{minipage}[b]{0.4\textwidth}
  \includegraphics[width=\textwidth]{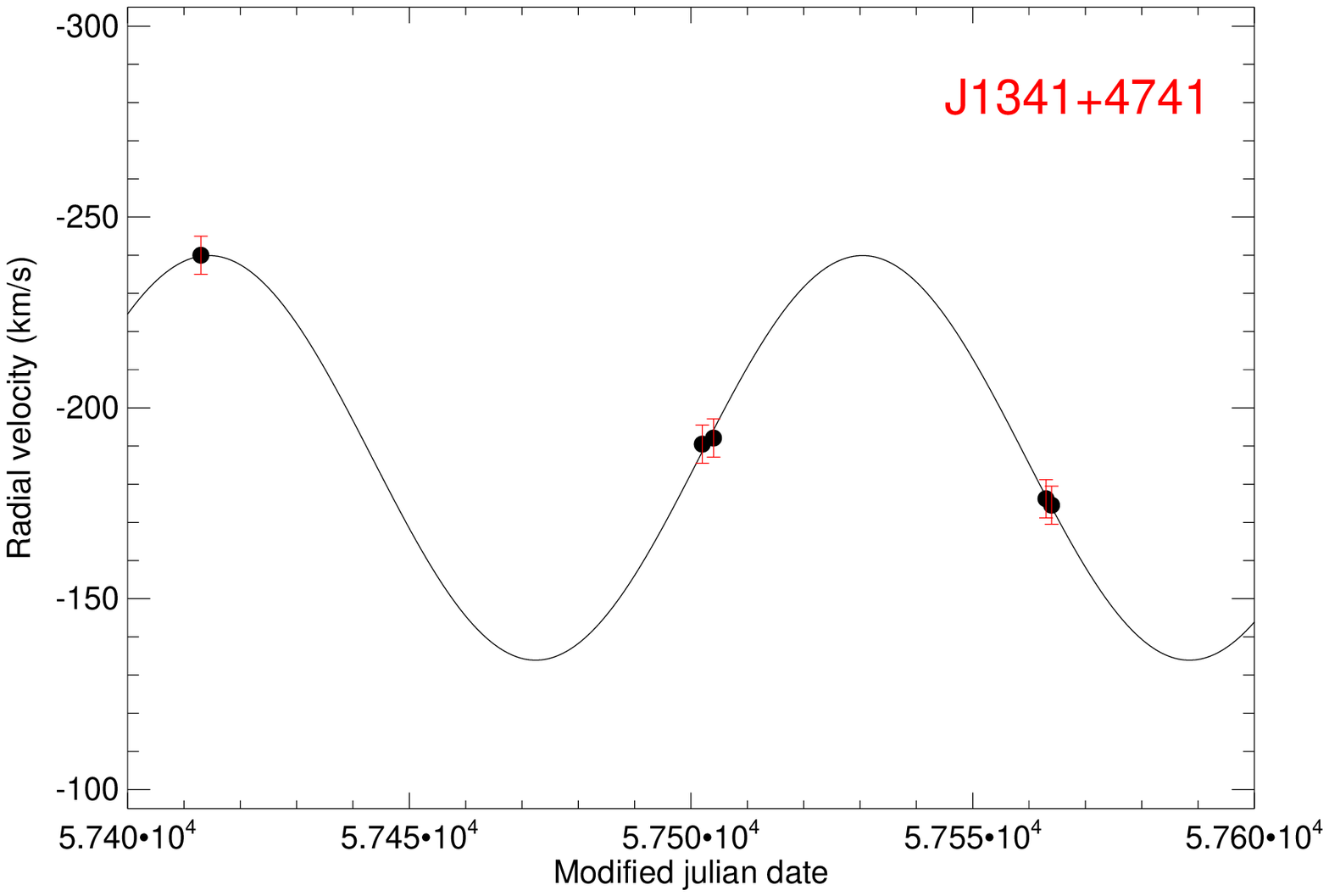}
  \caption{\normalfont \small Variation of radial velocity for \sdssone. The derived period is
116.0 days}
  \end{minipage}
\end{figure}

\section{STELLAR PARAMETERS}
Both photometric and spectroscopic data were used to derive estimates of
the stellar parameters for our program stars. The effective temperatures
were determined using various photometric observations in the literature
and the $T_{\rm eff}$-color relations derived by \citet{ramirez}. They
were found to be in close proximity (a difference of $~$40\,K was found)
to values obtained using \citet{alonso1996} and \citet{alonso1999}. The
$V-K$ temperature estimate is expected to be superior, as it is least
affected by metallicity and the possible presence of molecular carbon
bands. We also employed VOSA (http://svo2.cab.inta-csic.es/), the online
SED fitter \citep{bayo2008vosa}, to derive the temperatures using all of
the available photometry (optical, 2MASS, and WISE). A Bayesian fit
using the Kurucz ODFNEW/NOVER model was used to obtain the SED
temperature. Final fits for the two stars are shown in Figures 3 and 4.

$T_{\rm eff}$ estimates have also been derived spectroscopically, by
demanding that there be no trend of Fe~I line abundances with excitation
potential (shown in the upper panels of Figures 5 and 6), as well by
fitting the $H_{\alpha}$ profiles. Estimates for the effective
temperatures of our target stars are listed in Table 3.~ For
\sdssone, ~ we have adopted the temperature obtained from fitting of
the $H_{\alpha}$~ wings, as it is highly sensitive to small variations in
temperature. For \sdsszero, the $H_{\alpha}$~ profile was asymmetric, and
thus it could not be used for accurate measurement of temperature. So,
the temperature obtained from Fe~I line abundances was adopted.

Surface gravity, $\log (g)$, estimates for the two stars have been
determined by the usual technique that demands equality of the iron
abundances derived for the neutral (Fe~I) lines and singly ionized
(Fe~II) lines. We used 7 Fe~II lines and 82 Fe~I lines for \sdsszero, and
5 Fe~II lines and 49 Fe~I lines for \sdssone; best-fit models for our
target stars are shown in the upper panels of Figures 5 and 6. The wings
of the Mg~I lines have also been fitted to obtain estimates for $\log
(g)$; best-fit models are shown in Figure 7.

The microturbulent velocity ($\xi$) estimates for each star have been
derived iteratively in this process, by demanding no trend of
Fe~I abundances with the reduced equivalent widths, and are plotted in 
the lower panels of figures 5 and 6. The
final adopted stellar atmospheric parameters are listed in Table 4.

\begin{table}
\begin{center}
\caption{Estimates of Effective Temperature} \label{tbl-3}
\begin{tabular}{cccccrrrrrr}
\tableline\tableline
Method &  $T_{\rm eff}$ (K) \\
\tableline
&SDSS~J0826+6125  &SDSS~J1341+4741\\
\tableline
$V-K$ &4453 &5827\\
SED &4500 &5500\\
H$\alpha$ &4400 &5450\\
Fe~I/Fe~II  &4300 &5400\\
\tableline
\end{tabular}
\end{center}
\end{table}

\begin{table}
\begin{center}
\caption{Adopted Stellar Parameters} \label{tbl-4}
\begin{tabular}{crrrrrrrrrrr}
\tableline\tableline
Object &$T_{\rm eff}$ (K) & $\log (g)$ & $\xi$ &[Fe$/$H]\\
\tableline
SDSS~J0826+6125 &4300 &0.40 &1.80 &$-$3.10\\
SDSS~J1341+4741 &5450 &2.50 &1.80 &$-$3.20\\
\tableline
\end{tabular}
\end{center}
\end{table}

\begin{figure*}[h!]
\epsscale{.80}
\includegraphics{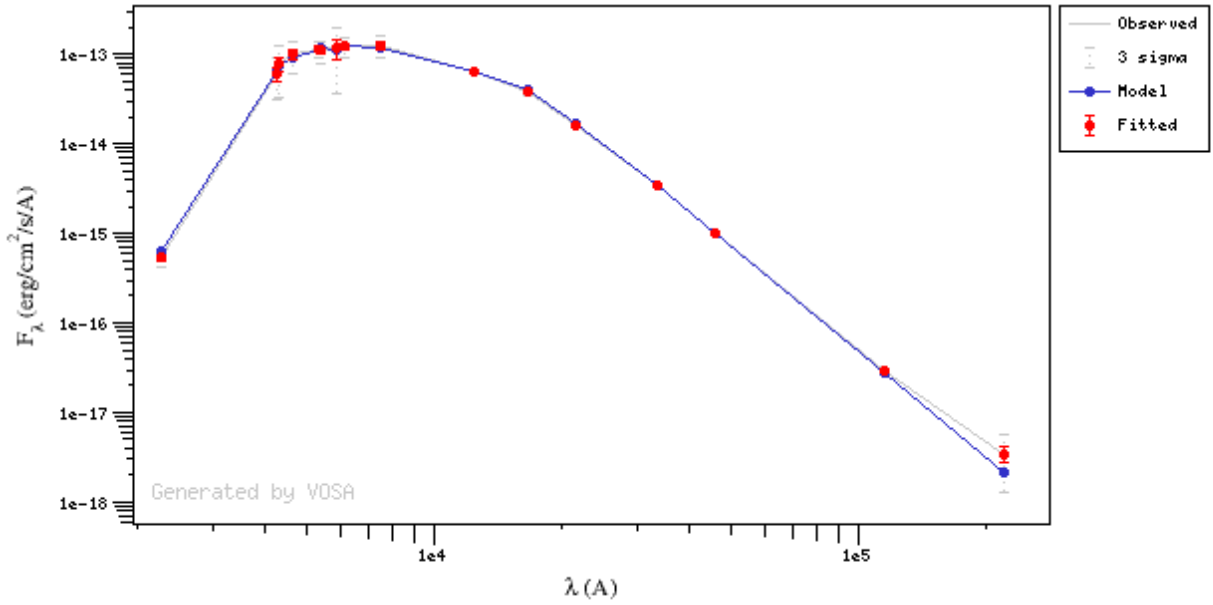}
\caption{ The SED obtained from VOSA for SDSS~J0826+6125 shows the
temperature to be $\sim$4500\,K.} \label{fig3}
\end{figure*}

\begin{figure*}[h!]
\epsscale{.80}
\includegraphics{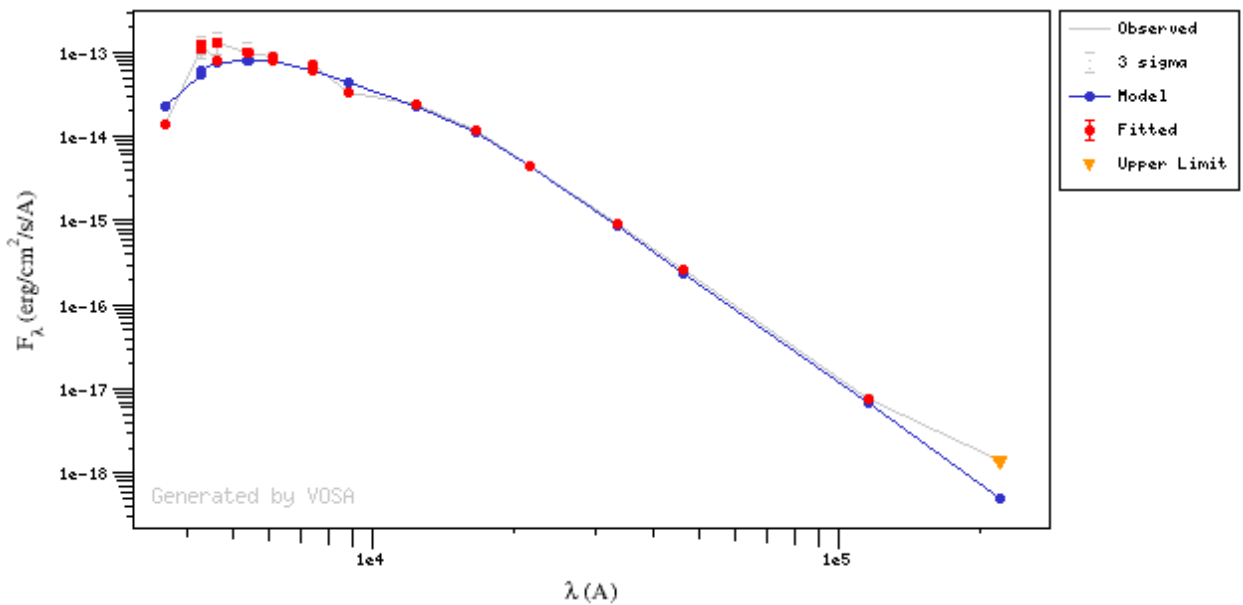}
\caption{The SED obtained from VOSA for SDSS~J1341+4741 shows
the temperature to be $\sim$5500\,K. } \label{fig4}
\end{figure*}

\begin{figure*}[h!]
\epsscale{.80}
\includegraphics{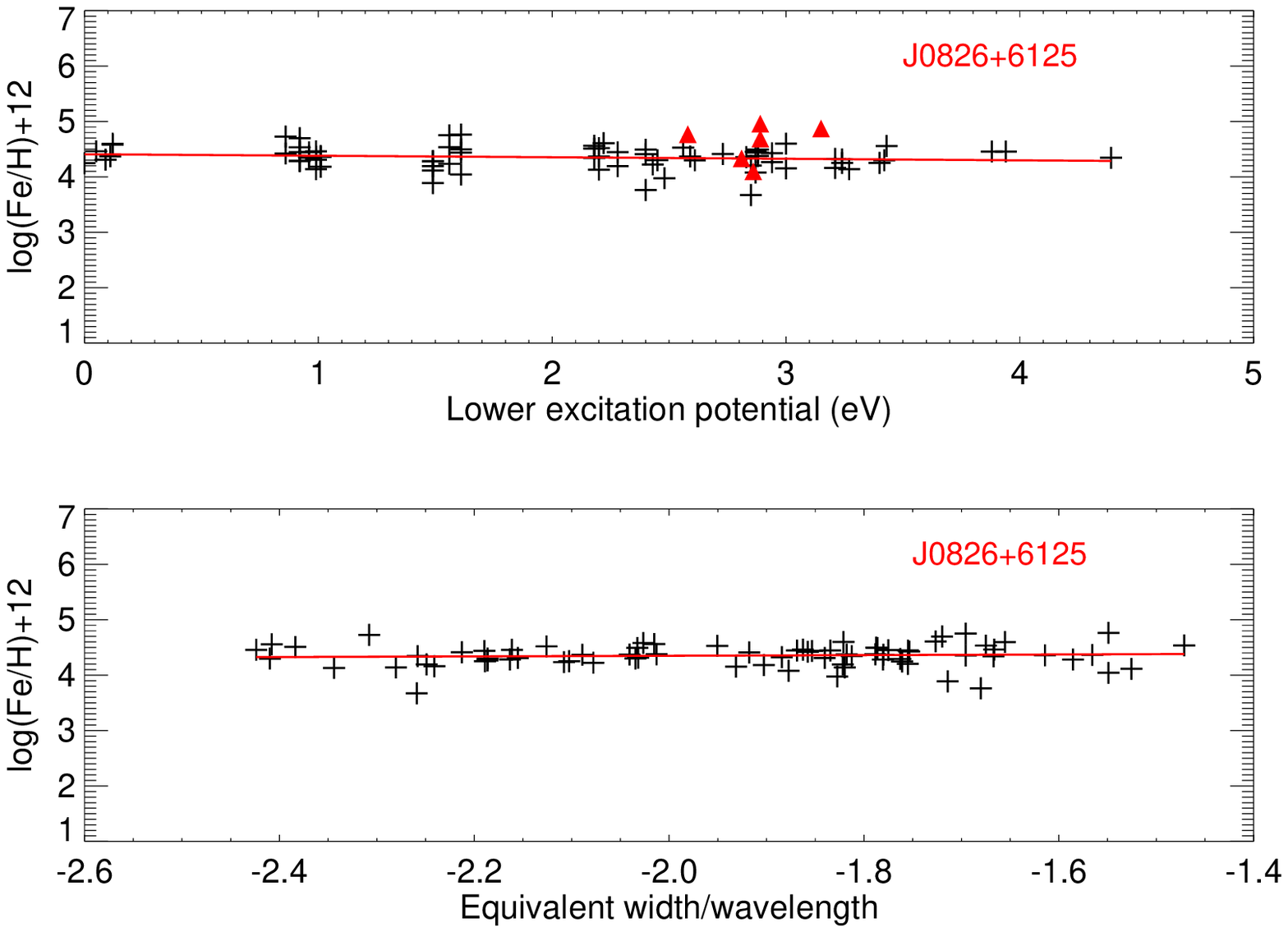}
\caption{ Top panel: Fe abundances derived from all lines, as a function of the
lower excitation potential, for the adopted model for
SDSS~J0826+6125. Lower panel: Fe abundances, as a function of
reduced equivalent widths, for the measured lines.} \label{fig5}
\end{figure*}

\begin{figure*}[h!]
\epsscale{.80}
\includegraphics{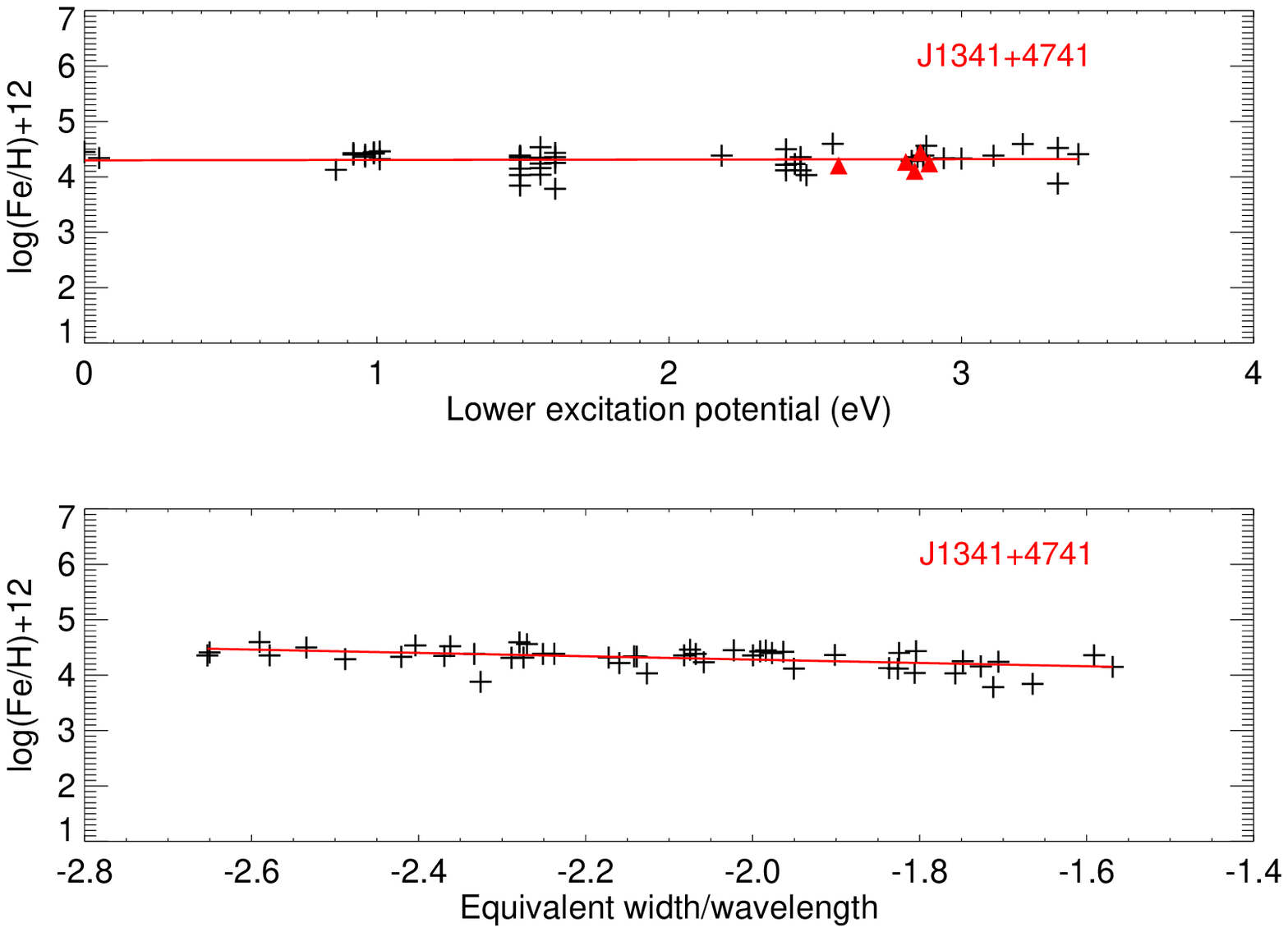}
\caption{Top panel: Fe abundances derived from all lines, as a function of the
lower excitation potential, for the adopted model for SDSS~J1341+4741.
 Lower panel: Fe abundances, as a function of reduced equivalent widths,
for the measured lines.} \label{fig6}
\end{figure*}

\begin{figure*}[h!]
\epsscale{.80}
\includegraphics{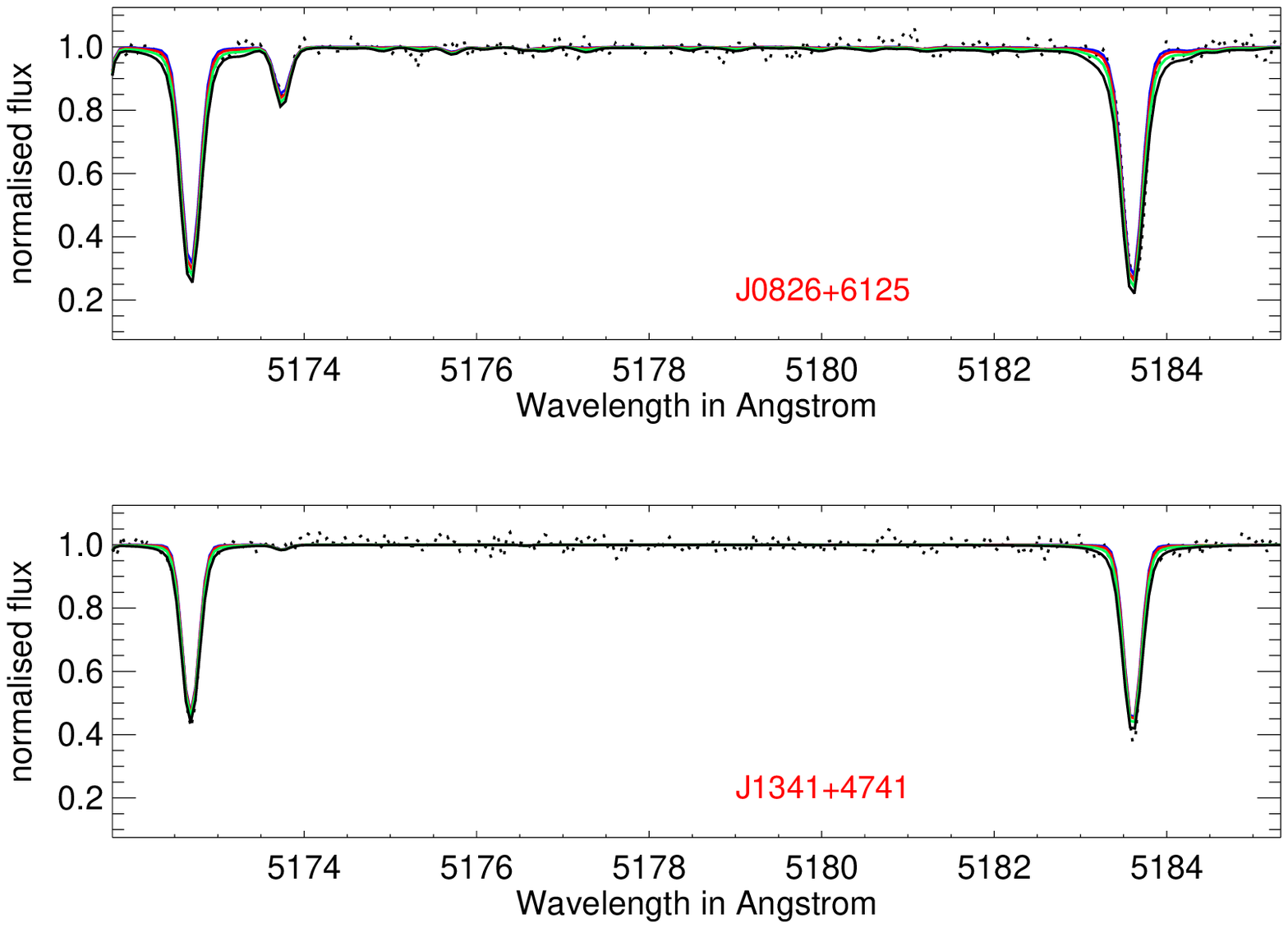}
\caption{High-resolution HESP spectra of SDSS~J0826+6125 (upper panel) and
SDSS~J1341+4741 (lower panel) in the region of the Mg~I triplet for
different values of $\log (g)$, in steps of 0.25 dex. The red solid line
indicates the best-fit synthetic spectrum. The adopted paremeters for
SDSS~J0826+6125 are $T_{\rm eff}$ = 4300~K and $\log (g) = 0.40$, while
those for SDSS~J1341+4741 are $T_{\rm eff}$ = 5450~K and $\log (g)$ =
2.50.} \label{fig7}
\end{figure*}

\subsection{Abundance Analysis}
To determine the abundance estimates for the various elements present in
our target stars we have employed one dimensional LTE stellar
atmospheric models (ATLAS9; \citealt{castellikurucz}) and the spectral
synthesis code turbospectrum \citep{alvarezplez1998}. We measured the
equivalent widths of the absorption lines present in the spectra, and
considered only those lines for abundance analysis whose equivalent
width is less than 120 m{\AA}, since they are on the linear part of the
curve of growth, and are relatively insensitive to the choice of
microturbulence. We measured the equivalent widths of 53 clean lines
present in the spectra of \sdsszero, among which 82 are Fe~I lines,
and 122 clean lines for \sdssone, among which 49 are Fe~I lines. We
have adopted the solar abundances for each element from
\cite{asplund2009,scott2,scott1,scott3}; solar isotopic fractions were used for all the
elements. Version 12 of the turbospectrum code for spectrum synthesis
and abundance estimates have been used for the analysis.We have adopted
the hyperfine splitting provided by \cite{mcwilliam1998} and solar
isotopic ratios. We have also used 2D MARCS models~(\citealt{marcs2008})
to derive the abundances, but no significant deviations were obtained.
The abundances differed by a values ranging from 0.01 to 0.02 dex for
individual species.
\begin{table}
\begin{center}
\caption{Elemental Abundance Determinations for \sdsszero} \label{tbl-5}
\begin{tabular}{crrrrrrrrrrr}
\tableline\tableline
Elements &Species & $N_{lines}$ & A(X) & Solar & [X/H] & [X/Fe] & $\sigma$\tablenotemark{*} \\
\tableline

C\tablenotemark{s} &CH & \dots    &4.60 &8.43   &$-$3.92 &$-$0.82 &0.04 \\
N\tablenotemark{s} &CN & \dots    &6.00 &7.83   &$-$1.83 &+1.27 &0.03 \\
O\tablenotemark{s} &O I & \dots    &6.50 &8.69  &$-$2.19 &+0.91 &0.01 \\
Na\tablenotemark{s} &Na I &2 &3.30 &6.21        &$-$2.91 &+0.19\tablenotemark{b} &0.01 \\
Mg\tablenotemark{s} &Mg I &4 &5.05 &7.59        &$-$2.54 &+0.56 &0.01\\
Al\tablenotemark{s} &Al I &1 &3.40 &6.43        &$-$3.03 &+0.07\tablenotemark{b} &0.02 \\
Ca &Ca I &8 &3.68 &6.32        &$-$2.64 &+0.46 &0.06\\
Sc\tablenotemark{s} &Sc II &5 &$-$0.06 &3.15    &$-$3.21 &$-$0.11 &0.01\\
Ti &Ti I &7 &1.96 &4.93        &$-$2.97 &+0.13 &0.03\\
   &Ti II &6 &2.06 &4.93       &$-$2.87 &+0.23 &0.04\\
Cr &Cr I &3 &2.10 &5.62        &$-$3.52 &$-$0.42 &0.05\\
   &Cr II &2 &2.35 &5.62       &$-$3.27 &$-$0.17 &0.05\\
Mn\tablenotemark{s} &Mn I &4 &1.60 &5.42        &$-$3.82 &$-$0.72 &0.02\\
Co\tablenotemark{s} &Co I &2 &2.00 &4.93        &$-$2.93 &+0.17 &0.01\\
Ni &Ni I &3 &3.00 &6.20        &$-$3.20 &$-$0.10 &0.04\\
Zn &Zn I &2 &1.50 &4.56        &$-$2.96 &+0.14 &0.05\\
Sr\tablenotemark{s} &Sr II &2 &$-$0.90 &2.83    &$-$3.73 &$-$0.63 &0.01 \\
Y\tablenotemark{s} &Y II &1 &$-$1.47 &2.21      &$-$3.68 &$-$0.58 &0.01\\
Zr\tablenotemark{s} &Zr II &2 &$-$0.75 &2.59    &$-$3.34 &$-$0.24 &0.01\\
Ba\tablenotemark{s} &Ba II &2 &$-$1.80 &2.25    &$-$4.05 &$-$0.95 &0.01 \\

\tableline
\end{tabular}
\end{center}
\begin{tablenotemark}
    \newline
    $\sigma$\tablenotemark{*} indicates the error.
    \newline
    \tablenotemark{b} Values obtained after applying NLTE corrections.
    \newline
    \tablenotemark{s} Indicates abundances obtained using synthesis.
    \end{tablenotemark}
\end{table}

\begin{table}
\begin{center}
\caption{Elemental Abundance Determinations for \sdssone} \label{tbl-6}
\begin{tabular}{crrrrrrrrrrr}
\tableline\tableline
Elements &Species & $N_{lines}$ & A(X) & Solar & [X/H] & [X/Fe] & $\sigma$\tablenotemark{*} \\
\tableline
Li\tablenotemark{s} &Li I &1 &1.95 & \dots   & \dots   &  \dots  &0.01 \\
C\tablenotemark{s} &CH & \dots   &6.22 &8.43             &$-$2.21 &+0.99 &0.04 \\
N\tablenotemark{s}{\textdagger} &CN & \dots  &7.00 &7.83 &$-$0.83 &+2.37 &0.05 \\
Na\tablenotemark{s} &Na I &2 &2.80 &6.21                 &$-$3.41 &$-$0.21\tablenotemark{b} &0.01 \\
Mg\tablenotemark{s} &Mg I &5 &5.10 &7.59                 &$-$2.49 &+0.71 &0.01\\
Al\tablenotemark{s} &Al I &1 &3.2 &6.43                  &$-$3.23 &-0.03\tablenotemark{b} &0.02 \\
Si &Si I &1 &5.33 &7.51                 &$-$2.18 &+1.02 &0.07\\
Ca &Ca I &11 &3.60 &6.32                &$-$2.72 &+0.48 &0.05\\
Sc\tablenotemark{s} &Sc II &3 &-0.1 &3.16                &$-$3.26 &$-$0.06 &0.01\\
Ti &Ti I &4 &2.23 &4.93                 &$-$2.70 &+0.50 &0.05\\
   &Ti II &13 &1.89 &4.93               &$-$3.04 &+0.16 &0.04\\
Cr &Cr I &6 &2.31 &5.62                 &$-$3.31 &$-$0.11 &0.04\\
   &Cr II &1 &2.77 &5.62                &$-$2.85 &+0.35 &0.06\\
Mn &Mn I &5 &1.89 &5.42                 &$-$3.53 &$-$0.33 &0.05\\
Co &Co I &2 &1.99 &4.93                 &$-$2.96 &+0.24 &0.05\\
Ni &Ni I &4 &3.35 &6.20                 &$-$2.85 &+0.35 &0.04\\
Sr\tablenotemark{s} &Sr II &2 &-0.88 &2.83               &$-$3.71 &$-$0.51 &0.01\\
Ba\tablenotemark{s} &Ba II &2 &-1.68 &2.25               &$-$3.93 &$-$0.73 &0.01 \\
\tableline
\end{tabular}
\end{center}
\begin{tablenotemark}
\newline
    \textdagger Only upper limits could be derived.
    \newline
    $\sigma$\tablenotemark{*} indicates the error.
    \newline
    \tablenotemark{b} Values obtained after applying NLTE corrections.
    \newline
    \tablenotemark{s} Indicates abundances obtained using synthesis
    \end{tablenotemark}
\end{table}

\section{ABUNDANCES}

\subsection{Carbon, Nitrogen, and Oxygen}
Carbon-abundance estimates for our stars were derived by iteratively
fitting the CH bandhead region with synthetic spectra, and adopting the
value that yields the best match. We have used the CH molecular line
list compiled by Bertrand Plez \citep{plez2005}. The CN and CH
molecular linelists are taken from the Kurucz database.

For \sdsszero, the O~I at 630 nm was used to measure the
oxygen abundance, which was found to be strongly enhanced, [O/Fe] =
+0.91. The chemical equilibrium of CO is taken into consideration in the
turbospectrum synthesis code \citep{laverny}. We also have CO spectra, and
though noisy, its oxygen abundance is coonsistent with the estimates from O~I. The carbon
abundance was obtained from the CH $G$-band region, which yielded a value
of [C/Fe] $= -0.82$. We have also checked the sensitivity of the CH band for
various O abundances, but no variation could be detected. The C$_{\rm 2}$
molecular band at 516.5 nm also could not be detected, which is
consistent with a low C abundance. We could also detect the bandhead in
the region of the CN band at 3884\, {\AA}, and obtain an enhancement in
nitrogen corresponding to a value of [N/Fe] = +1.27.

For \sdssone, the derived fit to the CH $G$-band yielded [C/Fe] = +0.99,
clear evidence for its enhancement. Using medium-resolution
spectroscopy from SDSS, \cite{fernandez} has previously reported a carbon
abundance ratio of [C/Fe] = +0.95. The O~I line at 630 nm is too weak to be detected,
hence no meaningful O abundance could be derived for this star. The
signal-to-noise ratio at the region of CN band is too low to confirm
enhancement in nitrogen for this star; so we could only obtain an upper
limit of [N/Fe] $< +2.37$.

Fits for in the region of the CH $G$-band are shown for both stars
in Figure 8.

\begin{figure*}[h!]
\epsscale{.80}
\includegraphics{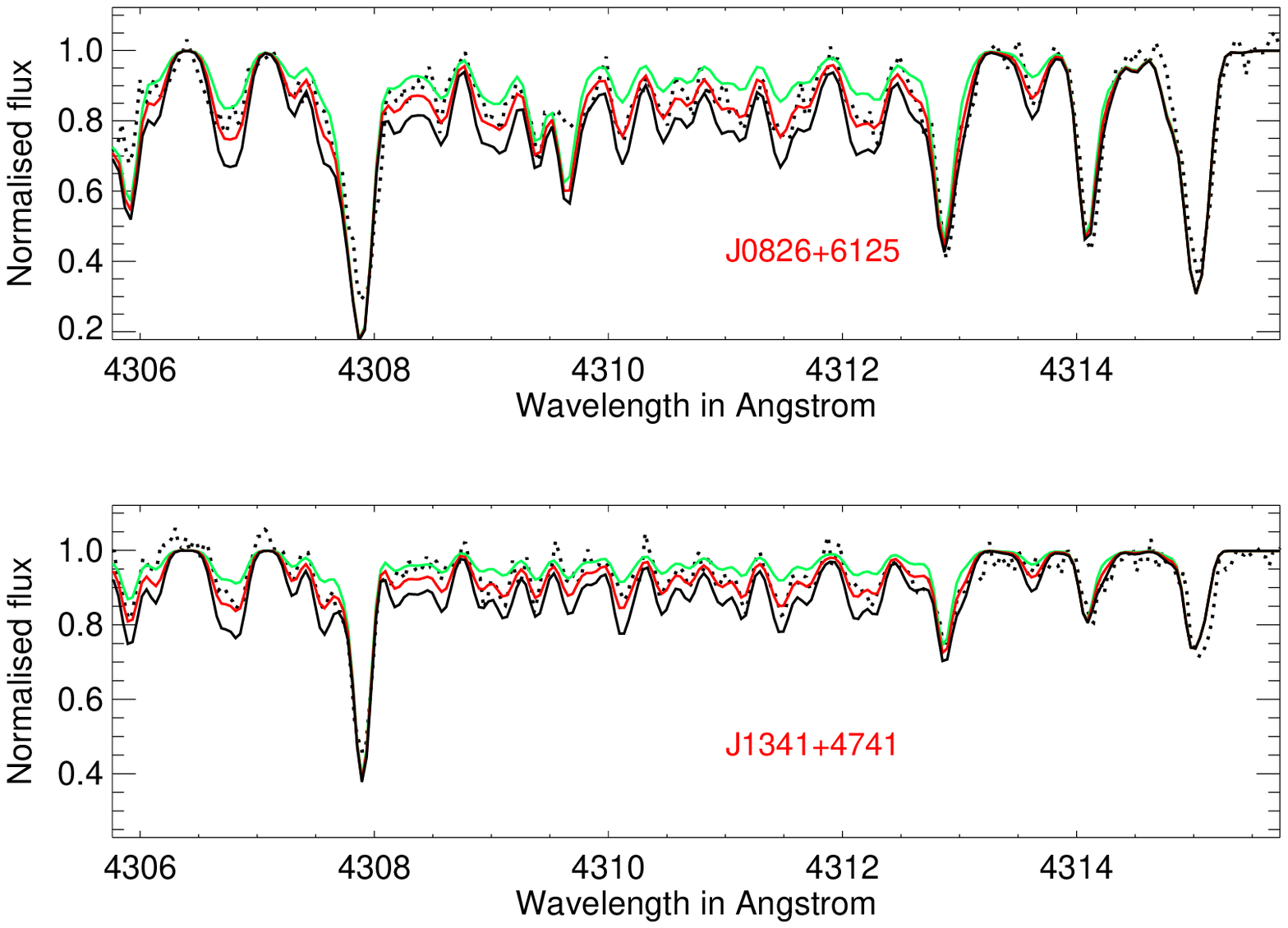}
\caption{High-resolution HESP spectra in the CH $G-$band region for
\sdsszero\ (upper panel) and \sdssone\ (lower panel).  The red solid line
indicates the synthetic spectrum corresponding to the best fit,
overplotted with two synthetic spectra with carbon 0.20 dex higher and
lower than the adopted value.} \label{fig8}
\end{figure*}

\subsection{The $\alpha$-Elements}
Several magnesium lines were detected in the spectra of our target stars.
Two of the lines in the Mg triplet at 5172\,{\AA}, and three other
lines at 4167\,{\AA}, 4702\,{\AA}, and 5528\,{\AA}, were used to obtain
the abundances. The derived [Mg/Fe] ratios for \sdsszero\ and \sdssone\ are
[Mg/Fe] = +0.56 and [Mg/Fe] = +0.71, respectively, values often found among
halo stars. The silicon lines at 5268\,{\AA} and 6237\,{\AA} were too
weak to be used for abundance estimates of \sdsszero, but for \sdssone, we
obtain [Si/Fe] = +1.0. It should be noted that Si may appear
over-abundant for metal-poor stars because LTE results are known to
overestimate the true value \citep{shi2012}.

Eight and eleven Ca~I lines were detected in the spectra of \sdsszero\ and
\sdssone, respectively, including the prominent lines at 4226.73\,{\AA}, 
4302.53\,{\AA}, and 4454.78\,{\AA}, and used to measured its abundance.
The measurements indicate slightly enhanced ratios of
[Ca/Fe] = +0.46 (for \sdsszero) and [Ca/Fe] = +0.48 (for \sdssone). The overall abundance of
the $\alpha$-elements is consistent with the typical 
halo enhancement of [$\alpha$/Fe] =  +0.4.

\subsection{The Odd-Z Elements}
The sodium abundance is determined from the Na~$D_{1}$ and $D_{2}$
resonance lines at 5890\,{\AA} and 5896\,{\AA}. The aluminium abundance
is obtained from one of the resonance lines at 3961.5\,{\AA}. This line
is not the ideal indicator, as it can have large departures from LTE, as
discussed by \cite{baumuller}, who found it to be as large as +0.6 dex.
\cite{gratton2001} showed that incorporation of these corrections
improves the agreement between the values of aluminum abundances
obtained from this line and the high-excitation infrared doublet at
8773\,{\AA}, in the case of globular cluster dwarfs. Hence, we have
applied this non-LTE correction to Al in our abundance table. Aluminum
is slightly enhanced for \sdsszero, while Na tracks the iron content
of the stars. The scandium content is also very similar to iron. 
Na and Al are produced by the Ne-Na and Mg-Al cycles in intermediate and 
massive stars during H-shell burning. 
Sodium and aluminum  in the two stars could be due to a well-mixed ISM,
and unlikely to have received direct contribution from intermediate-mass
or massive-star winds.

\subsection{The Iron-Peak Elements}
Iron abundances for \sdsszero\ were calculated using 82 Fe~I lines and 7 Fe
II~lines found in the spectra; a difference of 0.3 dex was noted between
the derived abundances. This difference between Fe~I and Fe~II is in
agreement with the NLTE effects explored by \cite{asplund2005}. Iron
abundances for \sdssone\ were calculated using 49 Fe~I lines and 7 Fe~II
lines found in the spectra; a difference of ~0.5 dex between the
abundance values obtained from these lines was found, which is rather
large.

We also detected the iron-peak elements Mn, Cr, Co, Ni, and Zn in our
target stars. Mn and Cr are products of incomplete explosive silicon
burning, and their abundances decrease with decreasing metallicity
\citep{mcwilliam1995, ryan1996, carretta2002}. For \sdsszero, the abundance
of Mn was derived from the resonance Mn triplet at 4030\,{\AA} and
three weaker lines near 4780\,{\AA}. Cr lines are measured from 4 lines,
including the stronger ones at 4646\, {\AA} and 5206\,{\AA}. Products of
complete silicon burning, such as Co, Ni, and Zn, have also been found in this
star; all of these elements are found to track the iron content.
For \sdssone, the abundance of Mn was derived from the resonance
Mn triplet at 4030\,{\AA} and an additional line at 3823\,{\AA}. The
observed abundances of Mn and Cr are similar to other EMP stars. The
abundance derived for Ni using the 4 lines of this element present in
the spectrum of \sdssone\ is clearly higher relative to iron, [Ni/Fe] =
+0.35.

\subsection{The Neutron-Capture Elements}
Strontium and barium are the two neutron-capture elements detected in
the spectra of \sdssone. Resonance lines of Sr~II at 4077\,{\AA}
and 4215\, {\AA} are detected in both of our target stars. \sdsszero\
is found to be under-abundant in both strontium and barium, with
abundances of [Sr/Fe] $ = -0.63$ and [Ba/Fe] $= -0.95$, respectively.
The other neutron-capture elements found in this star are Y and Zr,
which are under-abundant as well. \sdssone\ is also found to be
under-abundant in strontium compared to the solar ratio, 
[Sr/Fe] $ = -0.51$. Ba~II resonance lines at 4554\, {\AA} and 4937\,{\AA} were also
measured, and exhibited a considerable barium depletion, [Ba/Fe] $ =
-0.73$. Based on the clear under-abundance of the neutron-capture
elements, along with its strong carbon over-abundace, this star can be
confidently classified as a CEMP-no star. 

Best-fit spectra of the Sr and Ba syntheses for our two stars are shown
in Figures 9 and 10.

\begin{figure*}[h!]
\epsscale{.80}
\includegraphics{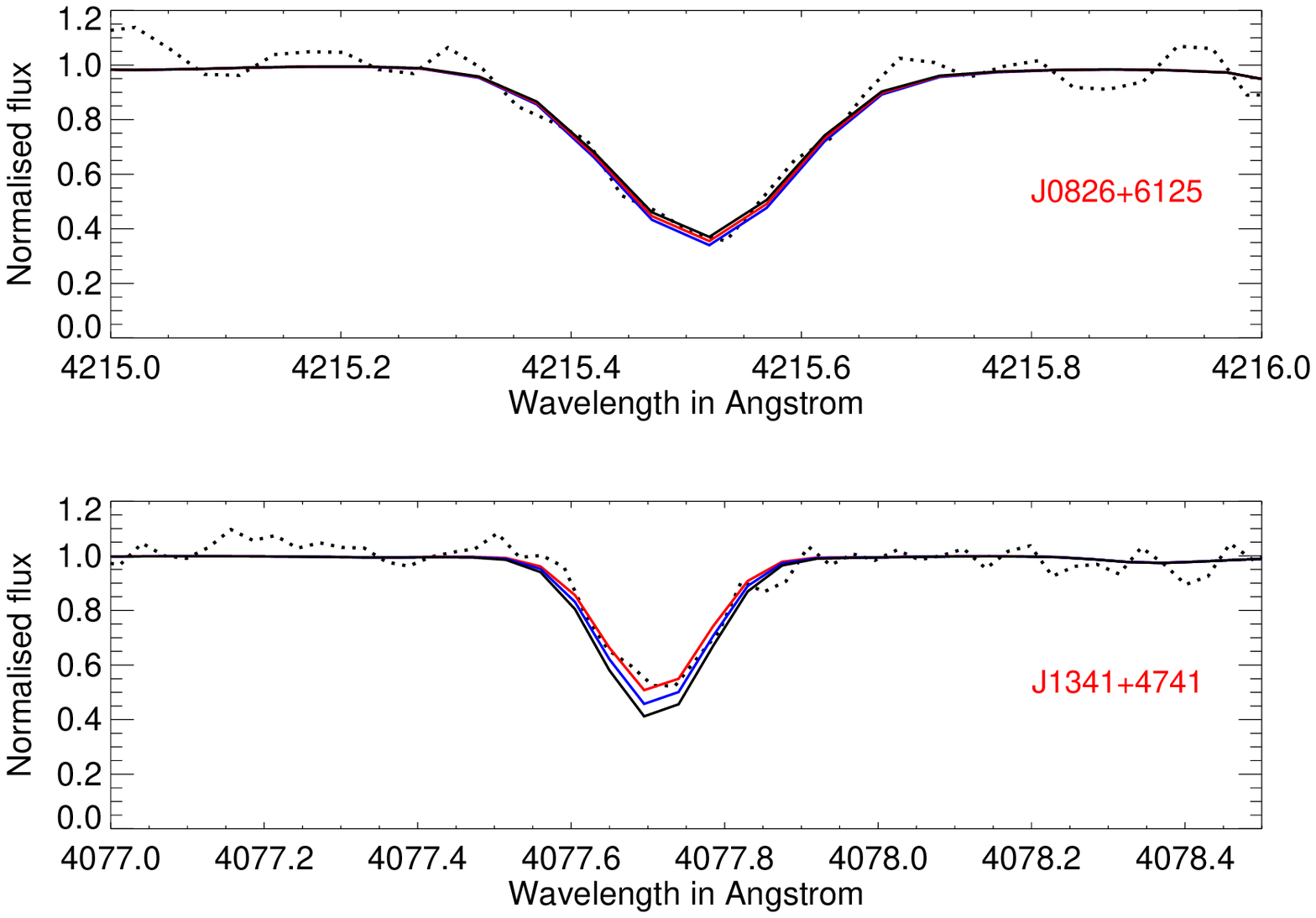}
\caption{Synthesis in the Sr II region for SDSS~J0826+6125 (upper panel)
and SDSS~J1341+4741 (lower panel). The red line indicates the best-fit,
overplotted with two synthetic spectra with Sr abundance 0.20 dex higher and
lower than the adopted value.} \label{fig9}
\end{figure*}

\begin{figure*}[h!]
\epsscale{.80}
\includegraphics{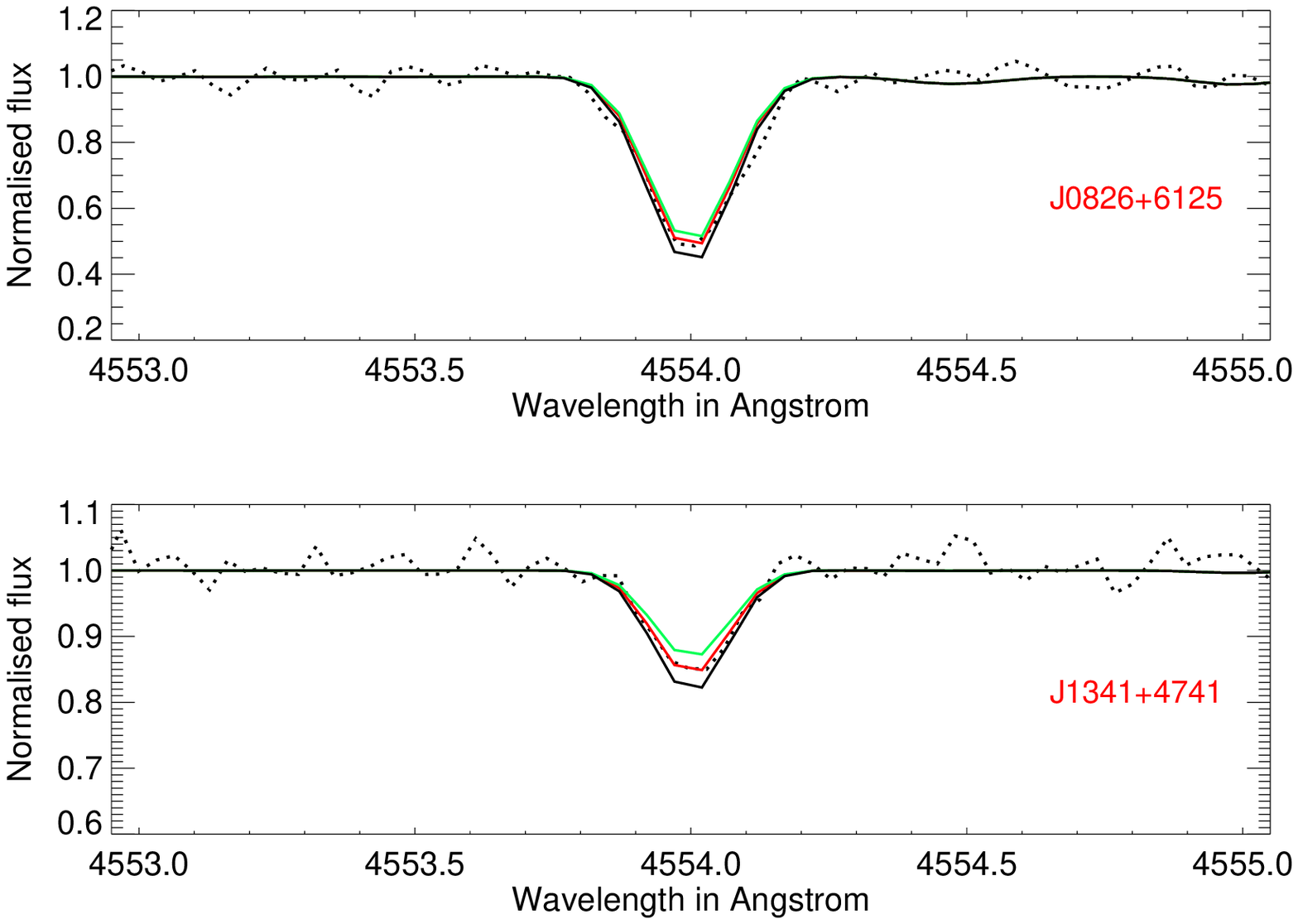}
\caption{Synthesis in the Ba II region for SDSS~J0826+6125 (upper panel)
and SDSS~J1341+4741 (lower panel). The red line indicates the best-fit,
overplotted with two synthetic spectra with Ba abundance 0.20 dex higher and
lower than the adopted value.} \label{fig10}
\end{figure*}

\subsection{Lithium}
Although lithium was not detected in \sdsszero, there is a strong feature
observed in \sdssone\ at 6707\,{\AA}, the Li doublet, from which we
obtain an abundance $A$(Li) = 1.95, which is similar to some other
CEMP-no stars (e.g , \citealt{sivarani2006}, \citealt{matsuno2017}). The
detection of lithium indicates that this star is unlikely to have
experienced AGB binary mass transfer or direct winds from a massive
star. Mass transfer from a low-mass AGB would produce large amounts of
carbon and deplete lithium, along with the production of $s$-process
enhanced material. A $(4-7M_{\odot})$ AGB star that had experienced hot bottom
burning produces large nitrogen and very low carbon. There are some
models in which AGB stars could produce lithium through the
Cameron-Fowler mechanism \citep{cameronfowler1971}. It is unclear if an
AGB with mass 3-4$M_{\odot}$ could explain the observed C, N, low $s$-process
elements, and lithium. Evolutionary mixing inside the star in its
subgiant phase might deplete the original lithium abundance of the
star-forming cloud. The synthesis for this element is shown in Figure
11.

\begin{figure*}[h!]
\epsscale{.80}
\includegraphics{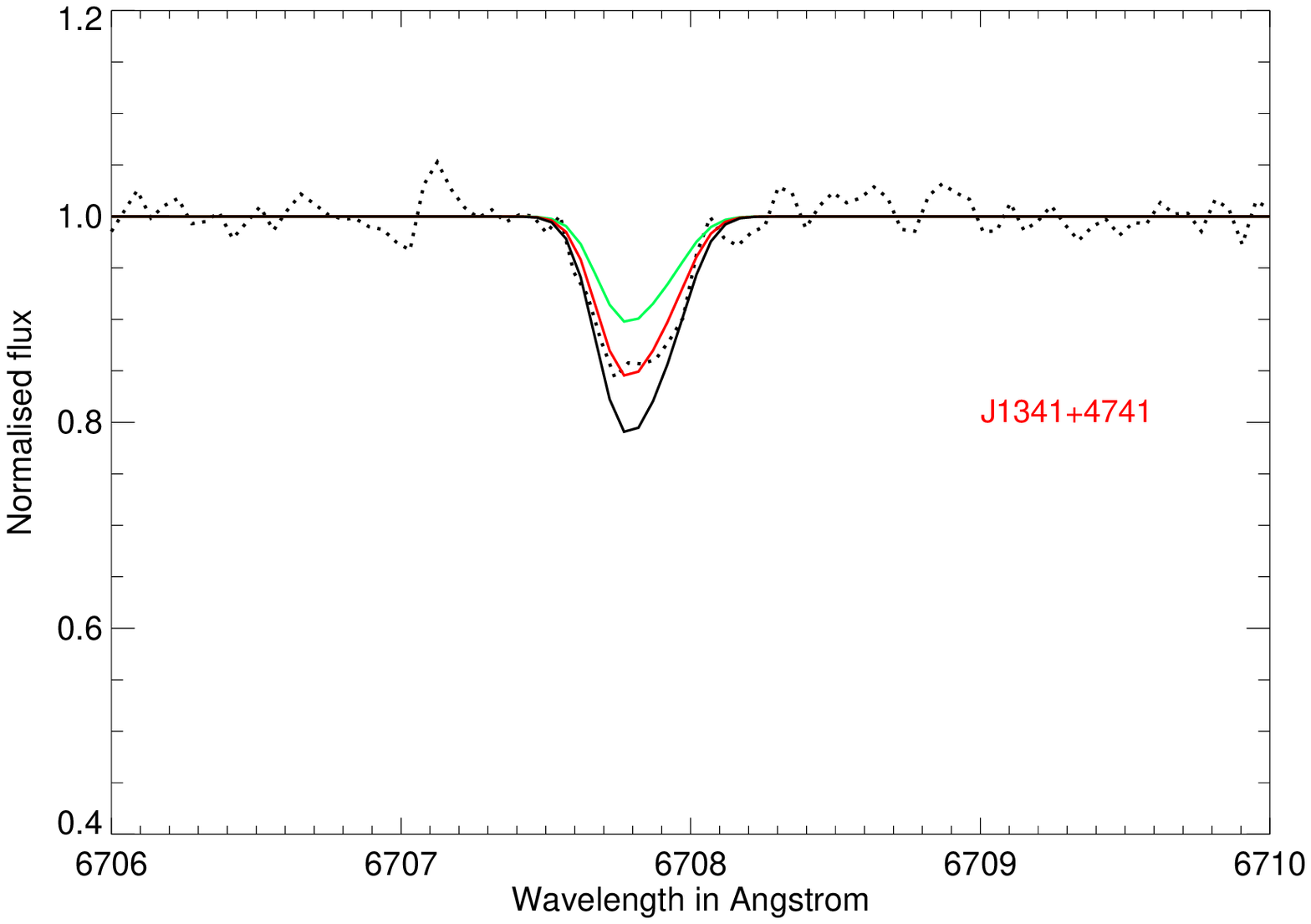}
\caption{Synthesis of lithium for SDSS~J1341+4741 at 
6707\,{\AA}.  The red line indicates the best-fit,
overplotted with two synthetic spectra with Li abundance 0.20 dex higher and
lower than the adopted value of A(Li) = 1.95.} \label{fig11}
\end{figure*}

\section{DISCUSSION}

\subsection{SDSS J082625.70+612515.10}

\subsubsection{Carbon, Nitrogen, and the Non-detection of Lithium}
In the {\it First Stars VI} paper, \citet{firststars6} found that carbon and
nitrogen were anti-correlated, and the faint halo stars
could be classified into two groups -- ``unmixed'' stars, which
exhibited C enhancement with N depletion, having $A$(Li) between 0.2 and
1.2, and ``mixed'' stars, which showed [C/Fe] $< 0.0$, [N/Fe] $> +0.5$,
and Li below the detection threshold. \sdsszero\ clearly falls into the
second group. Lithium is a very fragile element, which is destroyed at
temperatures in excess of 2.5 million K. Evidence for this can be seen
in previous samples of metal-poor stars; the $A$(Li) = 2.3 observed for
metal-poor dwarfs starts decreasing as the star ascends the giant
branch, to $A$(Li) $< 1.2$ for giants ({\it First Stars VII};
\citealt{firststars7}). The non-detection of lithium for this star could
be understood in this way.

In {\it First Stars IX}, \cite{firststars9} argued that such destruction
could be taken as a signature of mixing, and placed this mixed group of
stars higher up in the giant-branch stage of evolution. Other scenarios
for depletion of lithium, such as binary mass transfer, can be
eliminated for \sdsszero, as no such peculiar chemical imprints
have been found. During mixing, material from deeper layers where carbon
is converted to nitrogen is brought to the stellar surface. Figure 4 of
\cite{cayrel2004} shows the decline in the value of [C/Fe] for
temperatures below 4800~K in metal-poor stars, which is again attributed
to deep mixing at lower temperatures. With a $T_{\rm eff}$ of 4300~K and
a low $\log (g) = 0.4$, \sdsszero\ can be placed in the mixed group of
stars close to the tip of the red giant branch (RGB). Figure 12 shows the
position of the star in the $\log (g)$-$T_{\rm eff}$ plane, compared
with other metal-poor halo stars compiled in the SAGA database
\citep{sudasaga}. It sits right at the tip of the RGB.
Figure 13 compares the [C/N] ratio with metallicity of the halo stars
having carbon deficiency (and for which both estimates of carbon and
nitrogen are available). The abundance ratio of [C/N] for \sdsszero~ is
remakably low compared to other stars at the tip of the RGB.

\begin{figure*}[h!]
\epsscale{.80}
    \includegraphics[width=\textwidth]{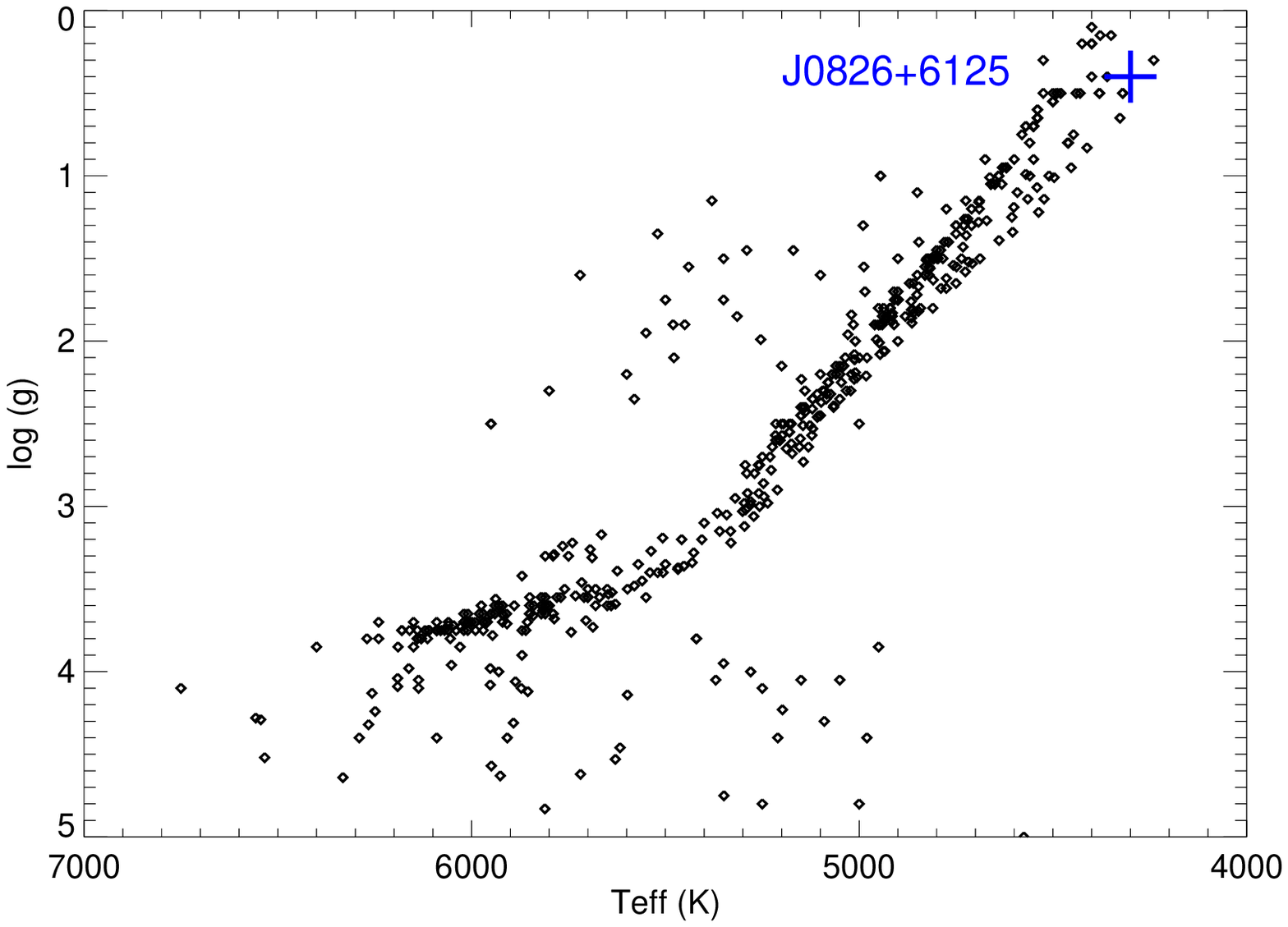}
    \caption{\normalfont \small The position of SDSS~J0826+6125 among other EMP halo stars in the log(g)-$T_{\rm eff}$ plane.
    The position of the star at the tip of the RGB is marked by the blue cross.} 
\end{figure*}

\begin{figure*}[h!]
\epsscale{.80}
  \includegraphics[width=\textwidth]{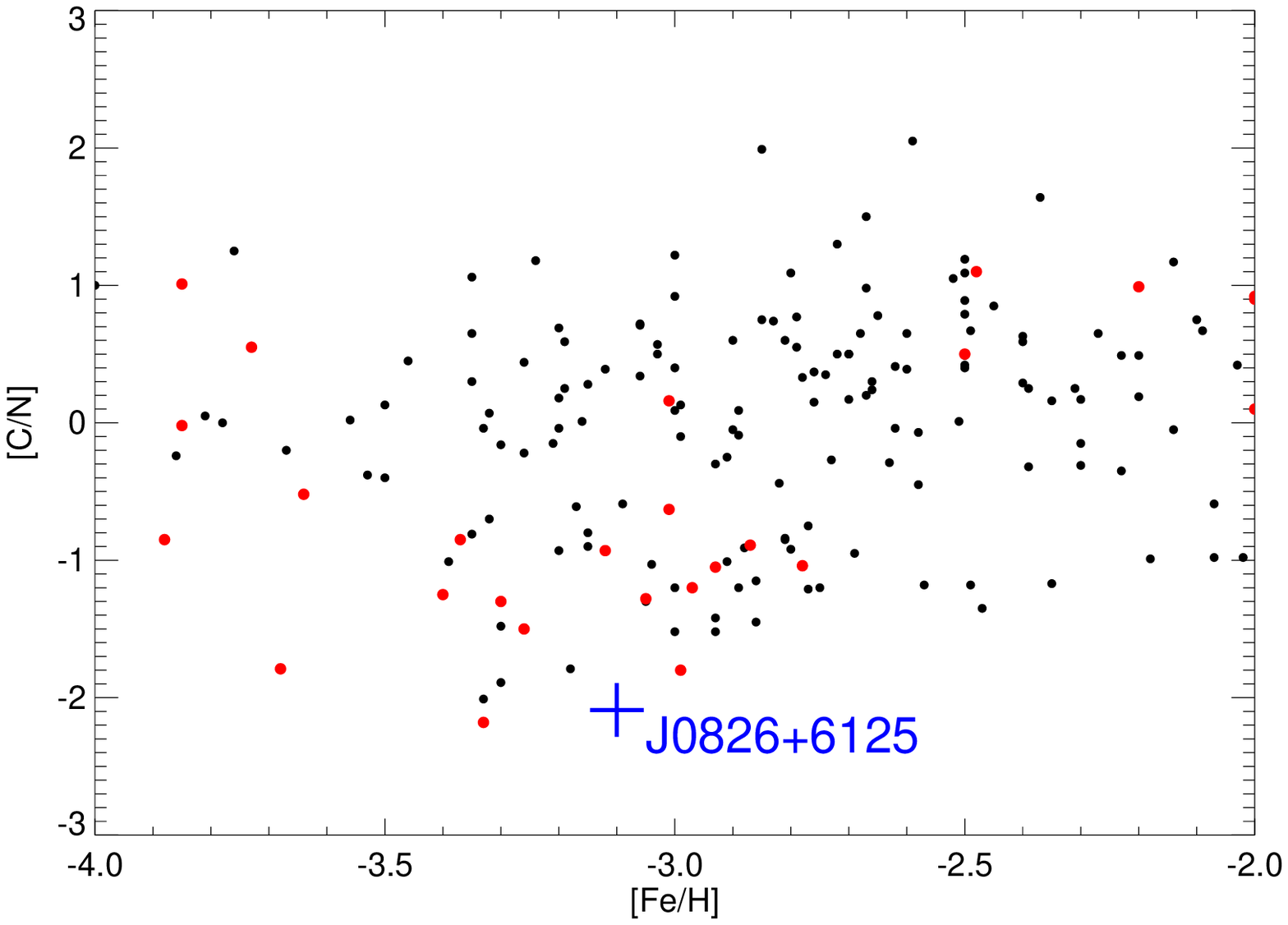}
  \caption{\normalfont \small The very low [C/N] abundance ratio other low-metallicity C-poor halo stars.
  SDSS~J0826+6125 is marked by the blue cross. The red dots mark the stars at the tip of the RGB with log(g)$<$ 1.}
\end{figure*}

\subsubsection{The Light Elements}
\sdsszero\ exhibits a low Na, high Mg, and low Al content, consistent
with the odd-even pattern expected to occur during massive-star
nucleosynthesis at low metallicities. A slight
enhancement of Na is observed, which could be an imprint of the previous
generations of stars which underwent the Ne-Na cycle, as it is not
possible to produce these elements in the RGB phase. Such an anomaly
could be similar to that seen in globular cluster stars
\citep{gratton2001,gratton2004}, which have undergone the AGB
phase and passed on processed material to a subsequent generation of
star formation in a closed system. Unfortunately, other signatures seen
in globular cluster stars, such as the C-N-O and O-Na-Mg-Al correlations
and anti-correlations \citep{shetrone1996,gratton2004,carretta2010,
coelho2011,meszaros2015} were not observed in this star.

\subsubsection{The Iron-Peak Elements}
Abundances of Fe-peak elements (Cr, Mn, Co, and Ni) for metal-poor stars
from the SAGA database are plotted, as a function
of metallicity, in Figure 14, along with the position of \sdsszero.
This star appears is relatively rich in Co, but poor in Cr, Mn, and Ni,
consistent with \citet{mcwilliam1995} and \citet{andouze}, who showed the
same trends for several stars with metallicity below [Fe/H] $= -2.4$.
The relative abundances of the Fe-peak nuclei could be well-explained
by their dependence on the mass cut of the progenitor supernova with
temperature, which gives rise to a photo-disintegration process 
\citep{woosleynweaver}. 

\begin{figure*}[h!]
\epsscale{.80}
\includegraphics{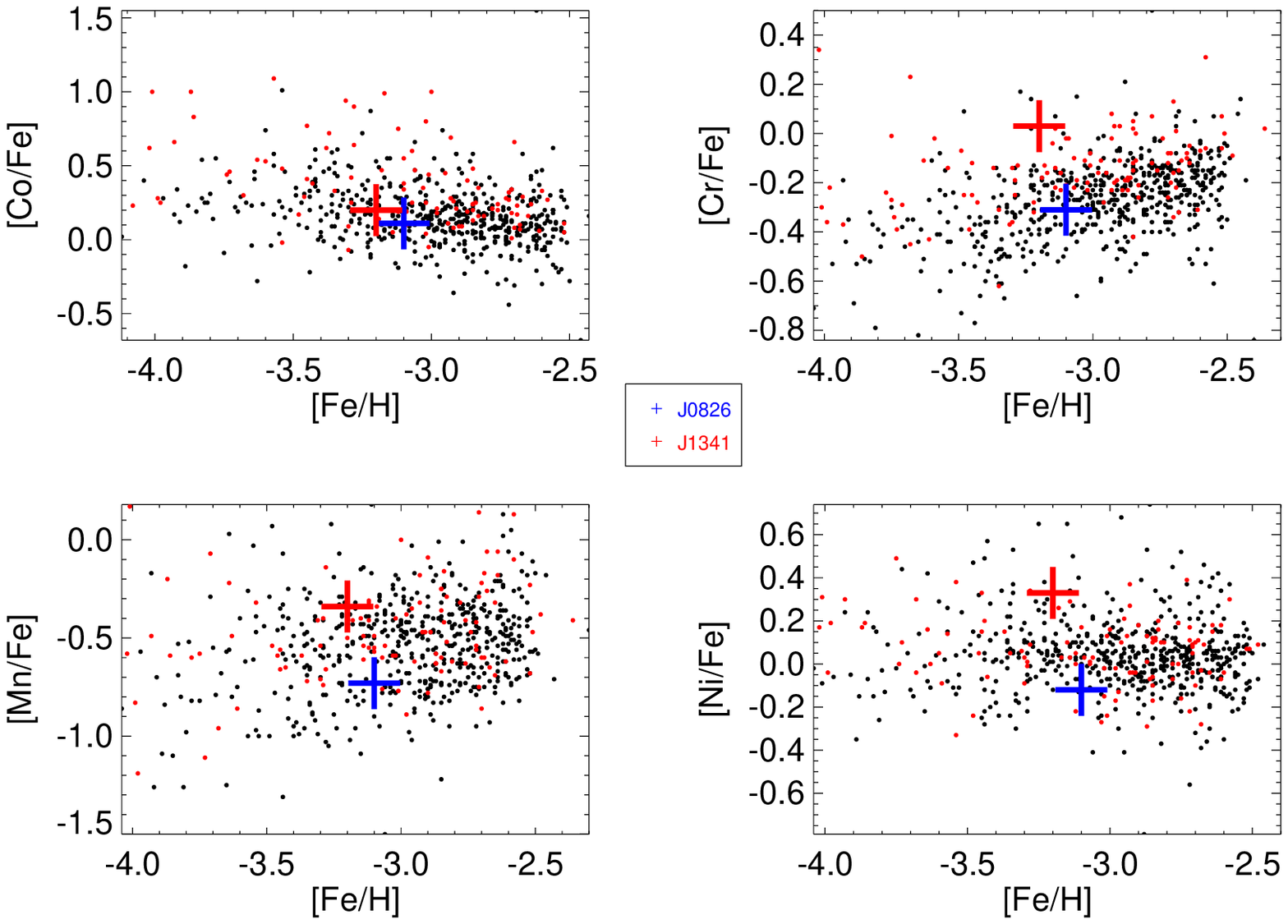}
\caption{Distribution of Fe-peak elements for Galactic halo stars. The
red dots represent the CEMP-no stars, while black dots represent C-normal halo stars.
The two program stars SDSS~J0826+6125 and SDSS~J1341+4741 are indicated
by blue and red crosses, respectively.} \label{fig14}
\end{figure*}

\subsubsection{The Neutron-Capture Elements}

Abundances of both the heavy and light $s$-process elements found in
\sdsszero\ are low, which is again consistent with the lack of available neutron flux
\citep{andouze}. The abundance values are very similar to other EMP
giants.

\subsubsection{The Asymmetric H$\alpha$ Profile of \sdsszero\ }

\sdsszero\  was observed several times, and an asymmetry in the H$\alpha$ 
profile was noted for all of the spectra. The profile also could not be
well-fit with synthetic spectra. The H$\alpha$ profile and its fit with
the model spectrum is shown in Figure 15. This could be due to the
inadequacy of the 1D stellar models, or it may be due to an extended
atmosphere present in the star. The H$\alpha$ profile was also found to
be not varying over several observation epochs indicating no ongoing
mass transfer. The extended atmosphere could be the result of past mass
transfer from an intermediate-mass AGB companion, or mixing due to first
dredge up of the star in the RGB phase. It is also possible that the
star itself is an AGB star (e.g., \citealt{masseron2006}).

\begin{figure*}[h!]
\epsscale{.80}
\includegraphics{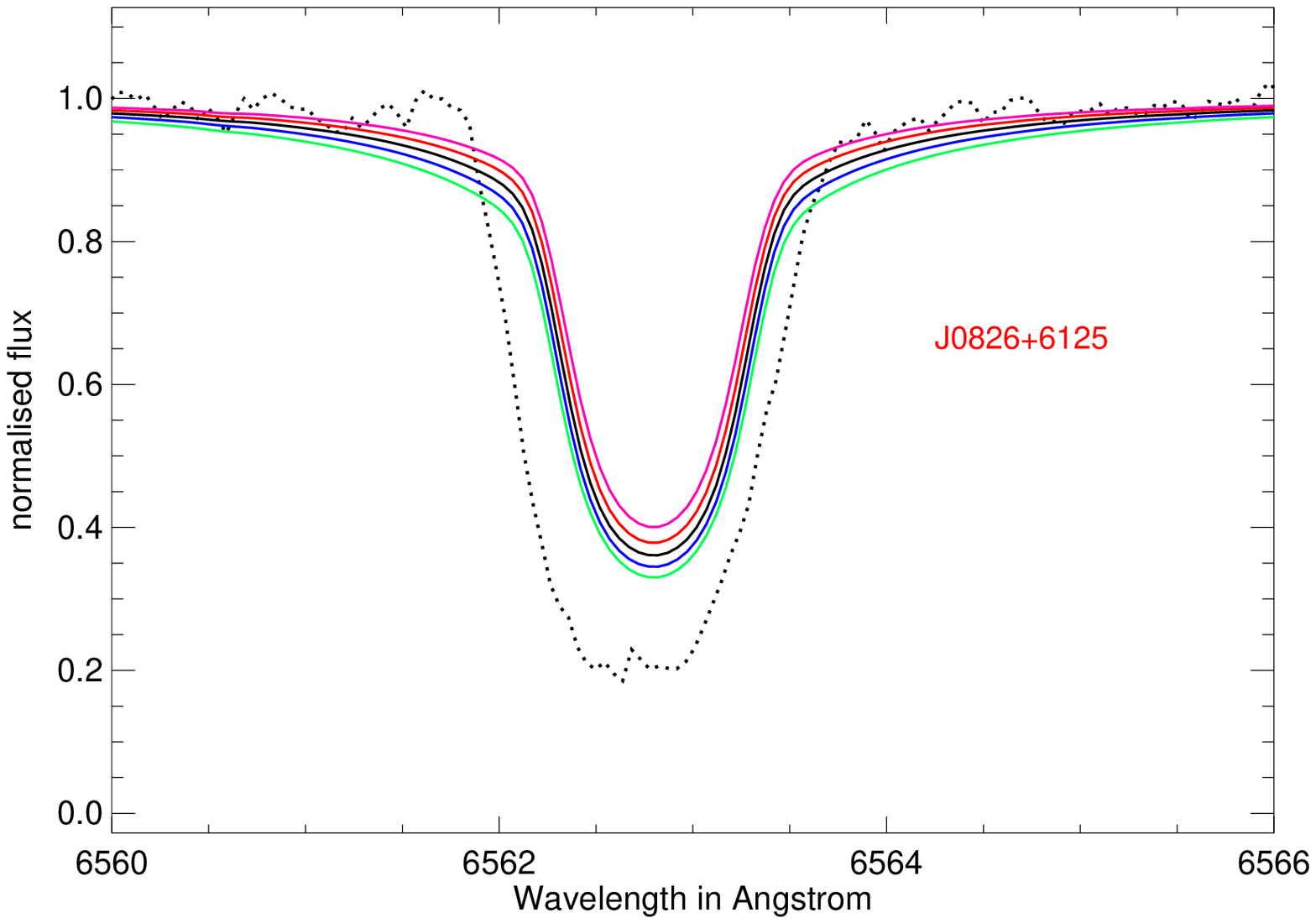}
\caption{The strange H$\alpha$ profile of \sdsszero, for different values
of temperature from 4200\,K to 4600\,K, in steps of 100\,K.} \label{fig15}
\end{figure*}

\subsection{SDSS J134144.60+474128.90}

\subsubsection{Lithium} 
We have obtained a measurement of $A$(Li) = 1.95 for \sdssone, which is 
lower than the Spite Plateau \citep{spitenspite} value of $A$(Li) =
2.2$\pm$ 0.1 \citep{Pinsonneault}, and much lower than the predicted
amount of Li from Big Bang nucleosynthesis ($A$(Li) = 2.75; \citealt{steigman2005}). Our
limited RV information for this star indicates clear variation, from
which we derive a possible period of 116 days. However, we have no other
evidence that a mass-transfer event may have occurred. The distribution
of lithium for CEMP-no stars, along with other EMP stars, is shown in
Figure 16. According to the analysis of \cite{meynet2010} and
\cite{masseron2012}, this star falls close to the edge of Li-depleted
stars ($A$(Li) = 2.00 is adopted as the separation between Li-normal and
Li-depleted metal-poor stars). A slight depletion from the Spite Plateau
value could be attributed to internal mixing of the star, or
the observed value of lithium for \sdssone\ may be the result of several
concurrent phenomena. 

\begin{itemize}

\item The ejected material from the progenitor SN will have depleted lithium 
abundance along with other nucleosynthetic elements and enhanced carbon
(for the case of \sdssone) that is mixed with the primordial cloud.
Depending upon the dilution factor in the natal cloud, it may be
possible to achieve the necessary lithium value \citep{piau2006,
meynet2010, maedermeynet2015}.  

\item A Spite Plateau value of Li was present in
the natal cloud of \sdssone, and it is depleted by thermohaline mixing or
meridional circulation \citep{masseron2012} in the star. If we consider
the current evolutionary state of the star to be in the RGB phase, this
could be a viable mechanism. 

\item Enhanced rotationally-induced mixing in
the RGB phase (following \citealt{denherwig}) can lead to formation of
lithium in the star, following depletion of all the primordial lithium.
It is very difficult to differentiate between an AGB or a massive
rotating star as the precursor using Li as the sole yardstick, as both
result in almost the same nucleosynthetic yield of Li
\citep{meynet2006,masseron2012}.
\end{itemize}

\begin{figure*}[h!]
\epsscale{.80}
\includegraphics{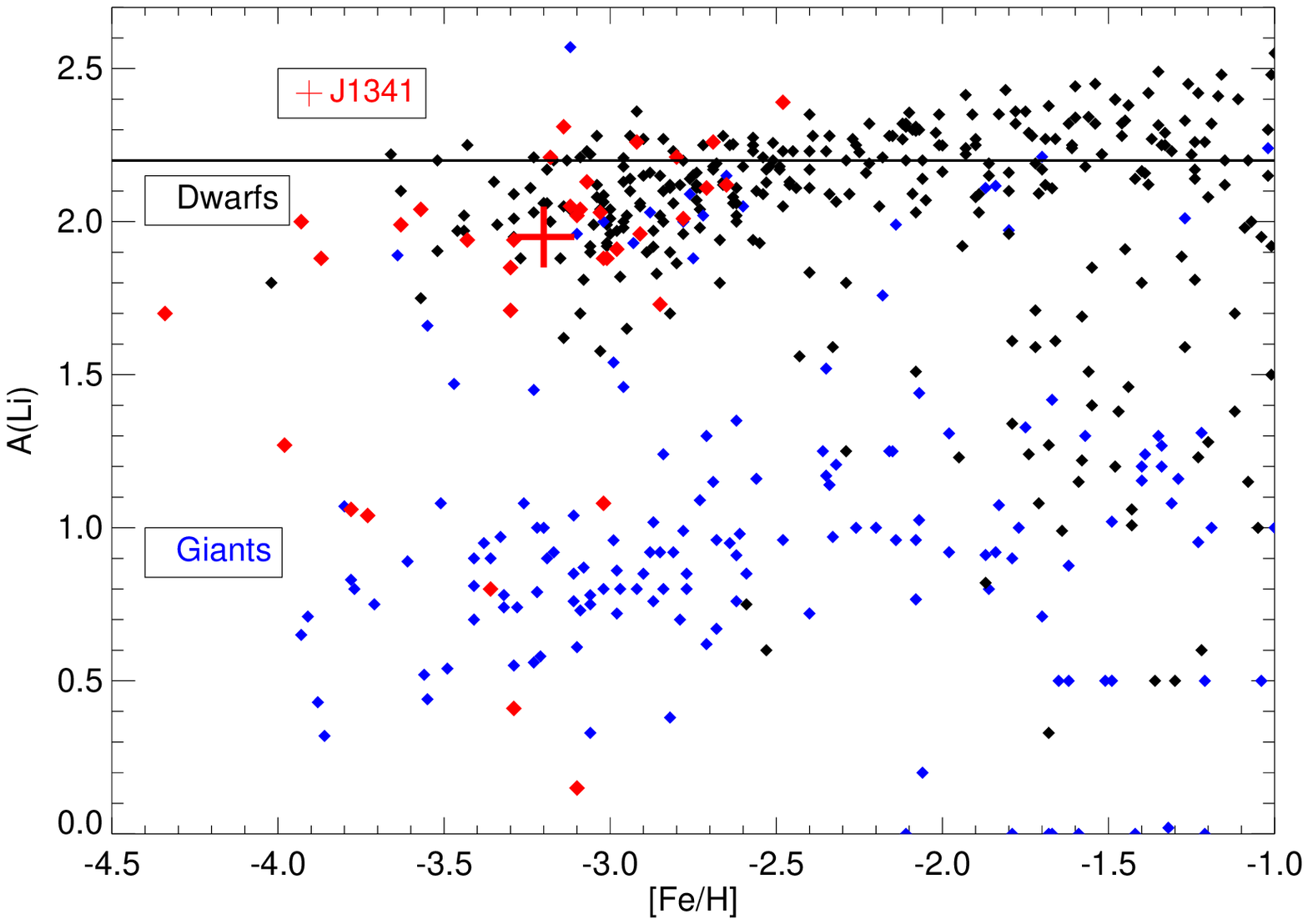}
\caption{Comparison of the observed lithium for CEMP-no stars, taken from the SAGA Database.
The blue dots mark the EMP giants while black dots are the EMP
dwarfs. Red points are the CEMP-no stars. The red cross marks the location of
SDSS~J1341+4741.} \label{fig16}
\end{figure*}

\subsubsection{Carbon}
According to \citet{spite2013} and \citet{bonifacio2015}, CEMP stars are
distributed along two bands in the $A$(C) vs. [Fe/H] plane. The upper
band is centered around $A$(C)$\sim 8.25$, and comprises relatively more
metal-rich CEMP-$s$ stars, while the lower band centered around $A$(C)
$\sim 6.50 $ comprises more metal-poor, and primarily CEMP-no, stars.
Further investigation by \cite{hansen2016b} also led the result that the
majority of the stars that are known binaries lie close to the upper
band. 

By expanding the list of CEMP stars with available high-resolution
spectroscopic analyses to include more evolved sub-giants and giants
(with the later giants having C abundances corrected for evolutionary
mixing effects; Placco et al. 2014), \citet{jinmiyoon} demonstrated that
the morphology of this abundance space is more complex, with three
prominent groups identified in the so-called Yoon-Beers diagram (their
Figure 1). They argued that a separation between CEMP-$s$ stars and
CEMP-no stars in their sample could be reasonably achieved by splitting
the sample at $A$(C) = 7.1, with the Group I CEMP-$s$ stars lying above
this level and the Group II and III CEMP-no stars lying below this
level. In this classification scheme, \sdssone, with $A$(C) $\sim
$6.22, can be comfortably identified as a Group II CEMP-no star. Hence,
the enhancement of carbon in this star is most likely to be intrinsic to
the star (i.e., the C was present in its natal gas), and not the
result of mass transfer from an extinct AGB companion. Thus, the
elemental-abundance pattern observed from this star is associated with
nucloesynthesis from a core collapse SN at early times, perhaps with
additional contributions from stars that formed and evolved within its
natal gas cloud.

\subsubsection{The Light Elements}
\sdssone\ exhibits the low [Na/Fe], high [Mg/Fe], and low [Al/Fe] ratios
expected from the odd-even pattern in massive-star nucleosynthesis
yields at low metallicities (e.g., \citealt{arnett1971};
\citealt{truran1971}; \citealt{peterson1976}; \citealt{umeda2000}; 
\citealt{hegerandwoosley2002}). The light elements closely follow the 
overall halo population observed in the Galaxy as well
\citep{cayrel2004}. Following \cite{jinmiyoon},
\sdssone is clearly a member of the Group II stars, and supports
a possible mixing and fallback SN as a likely progenitor.

\subsubsection{The Iron-Peak Elements}

Abundances of Fe-peak elements for \sdssone\ (Cr, Mn, Co, and Ni) are
shown in Figure 14, as a function of [Fe/H], compared with other CEMP-no
and C-normal EMP stars compiled from the SAGA database
\citep{sudasaga}. One feature that clearly stands out is the over-abundance
of Cr and Ni, and to some extent Mn. In the low-metallicity regime, the
stars are expected to show signatures of Type II SNe nucleosynthesis.
All three elements play key roles in determining the progenitor
population in the halo and the subsequent SNe yields. A decrease in
[Cr$/$Fe] and [Mn/Fe] with decreasing [Fe/H] should be accompanied with
enhancement in [Co/Fe], as a result of deeper mass cuts in the
progenitor SNe (refer to Figure 9 of \citealt{nakamura1999}). However,
enhancement in both [Cr/Fe] and [Mn/Fe] can be explained by an excess of
neutrons as well. Since neutron excess is a function of metallicity, we
have plotted [Cr/Fe] vs. [Mn/Fe] in Figure 17 to eliminate the trend
with Fe abundance (following, e.g., \citealt{carretta2002}). In this
plot, our program star occupies a relatively higher position amidst the
population of CEMP-no stars. From \citet{hegerandwoosley2002},
\citet{hegerandwoosley2008}, and \citet{qian2002}, it is known that very massive stars ($80 <
M/M_{\odot}< 240)$ belonging to Population III explode as
pair-instability SNe, which should not produce a correlation between
[Cr/Fe] and [Mn/Fe]. Thus, the presence of this correlation points us
towards Type II SNe associated with a relatively high-mass
($M/M_{\odot}< 80$), but not extremely high-mass, progenitor. 

Nickel is an extremely important element to gain further insight into
the nature of the progenitor of \sdssone. The depth of the gravitational
potential and amount of neutrino-absorbing material in the models are
the two factors that compete for the production of Ni in Type II SNe. In
very massive ($M/M_{\odot}> 30$) stars the deeper gravitational
potential restricts nickel from being ejected due to fallback, while
intermediate-mass ($10 < M/M_{\odot}< 20$) stars eject large amounts of
Ni because of a large neutrino-absorbing region
\citep{nakamura1999}. Thus, enhancement of Ni also points in the same
direction, that the progenitor is likely to be a massive ($20 <
M/M_{\odot}< 30$) star exploding as a Type II SNe in the early Galaxy. The
observations support the hypothesis of a mixing and fallback model
\citep{nomoto2013} with a lower degree of fallback, so as to eject a
larger mass of $^{56}$Ni.
\begin{figure*}[h!]
\epsscale{.80}
\includegraphics{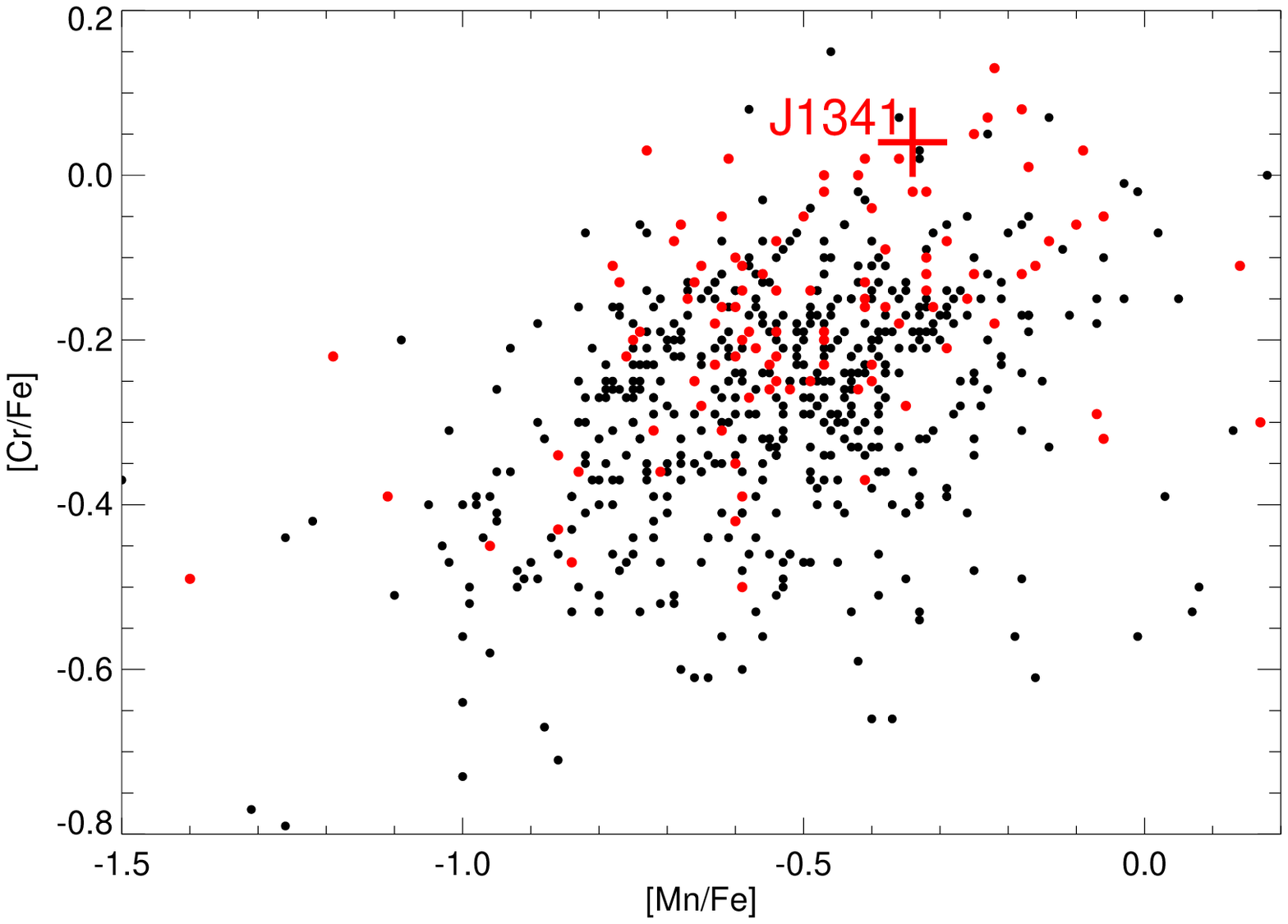}
\caption{The relative enhancement of Cr and Mn for
\sdssone, shown as a red cross, in the [Cr/Fe] vs. [Mn/Fe] space.
Red dots mark the CEMP-no stars while the black dots mark the EMP stars.} \label{fig17}
\end{figure*}

\subsubsection{The Neutron-Capture Elements}
The first $s$-process peak element Sr and the second $s$-process peak
element Ba have been detected in both \sdsszero\ and \sdssone, and they
exhibit under-abundances. The ratio of light to heavier neutron-capture
elements are sensitive to the nature of the progenitors. Neutron star
mergers are expected to produce heavy neutron-capture elements (e.g.,
\citealt{argast2004}) -- and have been observed to do so in the kilonova
SSS17a associated with GW170817 \citep{kilpatrick2017}, which
exhibited clear evidence for the presence of unstable isotopes created by the 
$r$-process \citep{drout2017,shappee2017}. SNe with jets (e.g.,
\citealt{winteler2012}; \citealt{nishimura2015}) may also produce heavy
neutron-capture elements. Formation of these systems may depend on the
environment as well.

\subsubsection{Nature of the Binary Companions of \sdsszero\ and \sdssone}
Both of the program stars exhibit clear RV variations,
indicating the likely presence of a binary companion. In the case of
\sdsszero, the enhanced abundances of N and under-abundance of
C indicates possible mixing of the atmosphere with CN-cycle products.
This can result from first dredge-up mixing in the star, which is
currently in the RGB, following mass transfer from an intermediate-mass AGB
star that might have gone through hot bottom burning \citep{lau2007,suda2012sf}. 
The low $\log (g)$ value of the star supports RGB mixing, although AGB
mass transfer cannot be ruled out. The non-detection of Li and peculiar
H$\alpha$ profiles could indicate either internal mixing or binary mass
transfer as well. In the case of an intermediate-mass AGB that goes
through hot bottom burning, the temperatures are sufficiently high for
the star to operate the CNO cycle. In that case, \sdsszero\ may be a 
true nitrogen-enhanced metal-poor (NEMP; see \citealt{johnson2007};
\citealt{pols2009}; \citealt {pols2012}) star, which are known to exist, but are relatively rare.

In the case of \sdssone, the binary companion did not likely contribute
through a mass-transfer event, since the Li abundance in the star is
similar to other EMP stars, although it is lower than the Spite Plateau
value. The mild depletion of Li could be due to binary-induced mixing or
internal mixing of the star during its sub-giant phase. It may well be
worthwhile to mount a RV-monitoring campaign for this and other
Li-depleted EMP stars to test for a possible binary-star origin to the
declining lithium abundance problem for stars with [Fe/H] $< -3.0$. 

\subsection{CEMP-no and EMP Stars}
From the above discussion, and based on previous studies, it is evident that
CEMP-no and C-normal EMP stars have very different origins. Even
within the sub-class of CEMP-no stars, there may well be different types
of progenitors. As discussed by \citet{jinmiyoon}, the Group II CEMP-no
stars could be associated with the faint mixing and fallback SNe,
whereas the Group III CEMP-no stars can be attributed to the spinstar
models, with a number of exceptions for both the classes
\citep{meynet2006,nomoto2013}. See also the discussion of the
progenitors for CEMP-no stars by \citet{placco2016}. Some of the CEMP-no
stars lying in the low $A$(C) region may have a binary component, but no
mass transfer is supposed to have taken place \citep{starkenburg2014,
bonifacio2015, jinmiyoon}, which is further strengthened by the only
``slight'' depletion of Li in \sdssone, as described in the previous
section.
Iron-peak elements can provide valuable insights regarding the
nucleosynthetic yields of their progenitor supernovae, as these elements
cannot be produced or modified during the post main-sequence
evolutionary stages of the star. Figure 14 shows the distribution of
some key Fe-peak elements for both CEMP-no and C-normal EMP stars.
Visual inspection suggests that Cr and Co are enhanced for the CEMP-no
population. We have compiled data from SAGA databse to see if there is
an enhancement of Cr in CEMP-no stars. The fit is given in Figure 18
for [Cr/Fe]. There is a slight offset between the EMP and CEMP-no stars,
but they exhibit similar increasing trends of [Cr/Fe] with [Fe/H]. We
have checked, and these behaviors apply to both dwarfs and giants. Similar offset could also be
noted for Co.
\citet{lai2008} and \citet{bonifacio2012} have considered the
discrepancies in the behavior of Cr between giants and dwarfs, since
Cr~II could be measured only in giants, while Cr~I is a resonance line,
and could suffer substantial NLTE effects. However, such issues are not
expected to play a substantial role when we compare only giants with
giants or dwarfs with dwarfs. Temperature and gravity do not play a
major role in deviations from LTE abundances \citep{bergemann}, 
so we have not used them to further refine our sample
from the archival data. 
Enhancement in [Cr/Fe] for CEMP-no stars with respect to C-normal EMP stars can play a
key role for understanding of the SNe ejecta and relevant mass cuts. It
would be very interesting to investigate the origin for this
discrepancy. 
\begin{figure*}[h!]
\epsscale{.80}
\includegraphics{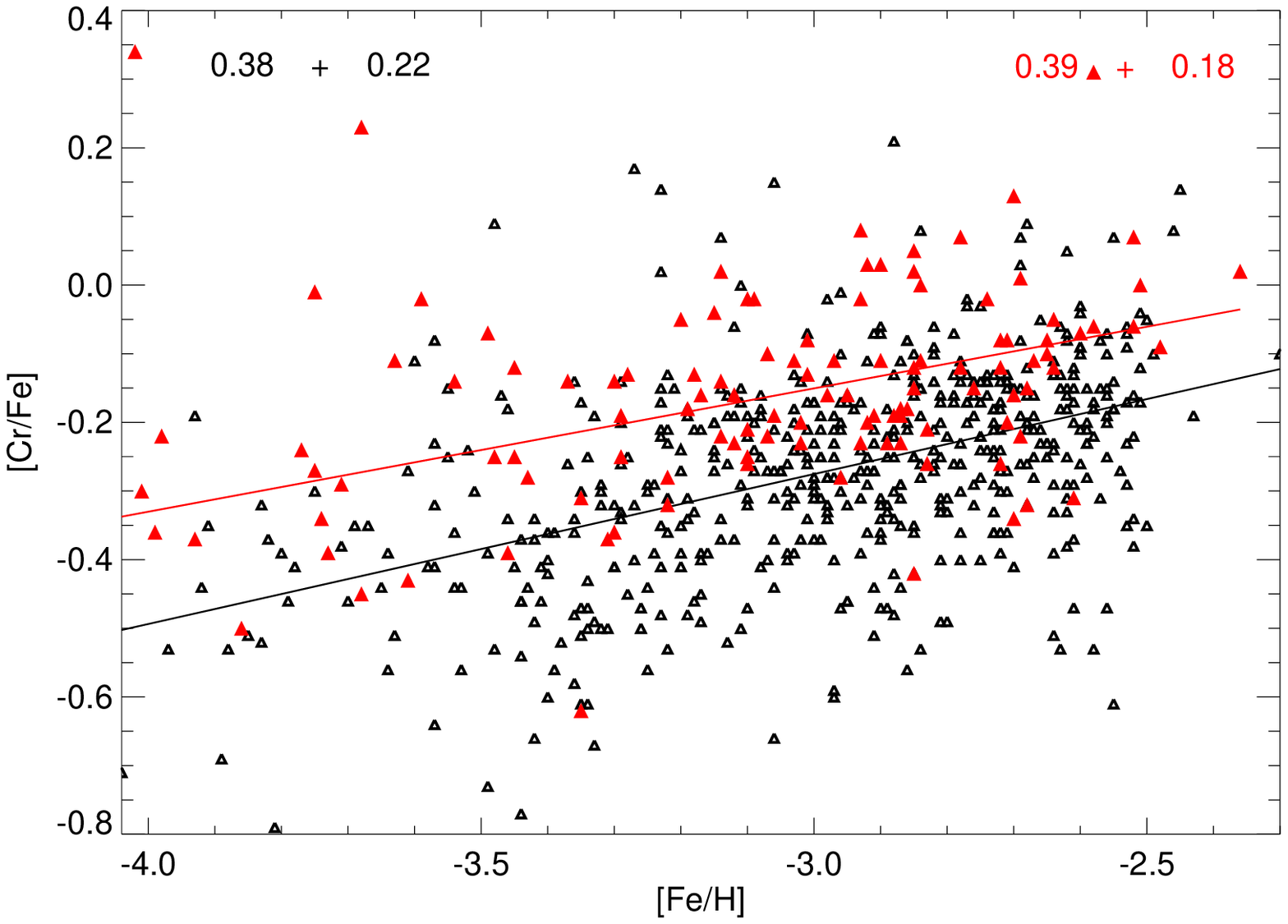}
\caption{Linear fit for
CEMP-no and C-normal EMP stars. Red is used for CEMP-no stars while
black is used for EMP stars. The slope and $\sigma$ are shown for each
fit in the corresponding color.} \label{fig18}
\end{figure*}

\section{Conclusion}
We have derived LTE abundances for \sdsszeroeight; it is mostly
consistent with behavior of other halo stars. The depletion in carbon
and enhancement in nitrogen could be due to internal mixing within the
star. It is unlikely that self enrichment similar to that is seen in
globular clusters has occured, due to the overabundance in oxygen. The
peculiar H$\alpha$ profiles of SDSS~J0826+6125 also supports the
possibility of mixing that might occur in an extended atmosphere. The
radial-velocity variation strongly suggests this star is a member of a
binary system, but it is likely there is no ongoing mass transfer, due
to non-variable peculiar H$\alpha$ profile over a period of year. 

\sdssonethree\ is a CEMP-no type star, and likely to be a member of
binary system. The lithium is detected and midly depleted, similar to
other EMP stars. Lithium in EMP dwarfs and CEMP-no stars exhibit similar
trends at different metallicities. Below [Fe/H] $< -3.0$, EMP and
CEMP-no stars often have lithium abundance below the Spite Plateau. We
also studied the trends of heavy elements among EMP and CEMP stars. At a
given metallicity, CEMP-no stars appear to have larger abundances of Cr.
This might provide important clues to the nature of the
progenitors that contributed to the origin of carbon.

\section{Acknowledgement}
We would like to take this opportunity to thank the HESP team for
succesful installation and maintenance of the high-resolution
spectroscope at the Hanle telescope, despite the numerous challenges and
hurdles that were encountered along the way. We would specially like to
thank Amit Kumar, M N Anand, Sriram and Kathiravan for their efforts. We
also extend our gratitude towards the observation team, namely Kiran B.S, Pramod Kumar,Lakshmi Prasad and
Venkatesh Shankar for their tireless effort. 

T.C.B. acknowledges partial support from grant PHY 14-30152 (Physics
Frontier Center/JINA-CEE), awarded by the U.S. National Science
Foundation (NSF). T.M. acknowledges support provided by the Spanish
Ministry of Economy and Competitiveness (MINECO) under grants
AYA−2014−58082-P and AYA-2017-88254-P.

\clearpage

\newpage
%\section{Appendix} 

\appendix

\clearpage

%\begin{center}
%\section*{Appendix}
%\end{center}

\begin{table}
\begin{center}
\caption{Linelist for \sdsszero} \label{tbl-1}
\begin{tabular}{ccccccccccccrrrrrr}
\tableline
Species &$\lambda$ &obs.eqw &error\tablenotemark{*} &Abundance \\
 \tableline
 CaI &4425.437 &34.000 &1.806 &3.284\\
 CaI &4435.679 &27.600 &1.628 &3.239\\
 CaI &5594.462 &30.300 &1.705 &3.574\\
 CaI &6122.217 &61.900 &2.437 &3.729\\
 CaI &6162.173 &82.000 &2.805 &3.867\\
 CaI &6439.075 &53.600 &2.268 &3.598\\
 CaI &6449.808 &8.000 &0.876 &3.870\\
 TiI &4981.731 &54.800 &2.293 &1.884\\
 TiI &4991.065 &52.700 &2.249 &1.970\\
 TiI &4999.503 &40.600 &1.974 &1.851\\
 TiI &5007.210 &52.100 &2.236 &2.188\\
 TiI &5014.187 &49.600 &2.182 &2.301\\
 TiI &5064.653 &30.200 &1.702 &1.804\\
 TiI &5210.385 &43.000 &2.031 &1.905\\
TiII &4470.857 &45.400 &2.087 &2.107\\
TiII &5129.152 &39.400 &1.945 &2.068\\
TiII &5154.070 &44.900 &2.076 &2.275\\
TiII &5185.913 &28.000 &1.639 &1.913\\
TiII &5188.680 &72.500 &2.638 &2.054\\
TiII &5381.015 &36.100 &1.861 &2.234\\
 CrI &4545.945 &7.700 &0.860 &1.986\\
 CrI &4646.148 &44.400 &2.064 &2.445\\
 CrI &5206.038 &81.300 &2.793 &2.170\\
 CrI &5298.277 &10.600 &1.009 &1.915\\
 CrI &5348.312 &6.700 &0.802 &1.849\\
CrII &4558.650 &8.500 &0.903 &2.249\\
CrII &4588.199 &11.400 &1.046 &2.574\\
\tableline
\end{tabular}
%\caption{jacob martin}
\end{center}
\end{table}

\begin{table}
%\scalebox{0.75}{
\begin{center}
%\caption{Linelist for \sdsszero} \label{tbl-1}
\begin{tabular}{ccccccccccccrrrrrr}
\tableline
Species &$\lambda$ &obs.eqw &error\tablenotemark{*} &Abundance \\
 \tableline
 FeI &4147.669 &68.747 &2.569 &4.283\\
 FeI &4174.913 &72.100 &2.631 &4.513\\
 FeI &4175.636 &23.000 &1.486 &3.837\\
 FeI &4202.029 &125.300 &3.468 &4.195\\
 FeI &4206.697 &90.700 &2.950 &4.492\\
 FeI &4216.220 &119.900 &3.392 &10.028\\
 FeI &4233.603 &63.000 &2.459 &3.989\\
 FeI &4250.787 &143.600 &3.712 &4.707\\
 FeI &4260.474 &89.000 &2.923 &3.780\\
 FeI &4415.123 &124.700 &3.460 &4.140\\
 FeI &4442.339 &73.700 &2.660 &4.481\\
 FeI &4447.717 &83.500 &2.831 &4.900\\
 FeI &4489.739 &99.300 &3.087 &4.696\\
 FeI &4528.614 &97.500 &3.059 &4.521\\
 FeI &4602.941 &69.399 &2.581 &4.353\\
 FeI &4733.592 &32.500 &1.766 &4.349\\
 FeI &4736.773 &27.200 &1.616 &4.294\\
 FeI &4871.318 &64.599 &2.490 &4.216\\
 FeI &4872.138 &68.300 &2.560 &4.464\\
 FeI &4890.755 &80.000 &2.771 &4.539\\
 FeI &4891.492 &85.983 &2.873 &4.212\\
 FeI &4903.310 &47.600 &2.137 &4.394\\
\end{tabular}
\end{center}
\end{table}

\begin{table}
\begin{center}
%\caption{Linelist for \sdsszero} \label{tbl-1}
\begin{tabular}{ccccccccccccrrrrrr}
\tableline
Species &$\lambda$ &obs.eqw &error\tablenotemark{*} &Abundance \\
 \tableline
 FeI &4918.994 &82.600 &2.816 &4.543\\
 FeI &4920.502 &99.200 &3.086 &4.563\\
 FeI &4939.687 &68.560 &2.565 &4.425\\
 FeI &4994.130 &95.200 &3.023 &4.709\\
 FeI &5001.864 &34.500 &1.820 &4.478\\
 FeI &5014.943 &18.900 &1.347 &4.499\\
 FeI &5049.820 &73.900 &2.663 &4.559\\
 FeI &5051.635 &106.799 &3.202 &4.689\\
 FeI &5068.766 &33.000 &1.780 &4.277\\
 FeI &5079.740 &69.700 &2.586 &4.504\\
 FeI &5083.339 &89.400 &2.929 &4.586\\
 FeI &5123.720 &64.100 &2.480 &4.332\\
 FeI &5127.359 &69.361 &2.580 &4.447\\
 FeI &5142.928 &79.200 &2.757 &4.544\\
 FeI &5150.840 &74.400 &2.672 &4.350\\
 FeI &5166.282 &89.600 &2.932 &4.306\\
 FeI &5171.596 &99.900 &3.096 &4.107\\
 FeI &5192.344 &78.394 &2.743 &4.607\\
 FeI &5194.942 &104.700 &3.170 &4.806\\
  FeI &5216.274 &85.100 &2.858 &4.588\\
 FeI &5225.526 &47.600 &2.137 &4.425\\
 FeI &5232.940 &92.300 &2.976 &4.439\\
 FeI &5247.050 &36.700 &1.877 &4.372\\
 FeI &5250.209 &49.400 &2.177 &4.593\\
 FeI &5250.646 &39.300 &1.942 &4.562\\
 FeI &5266.555 &61.700 &2.433 &4.287\\
 FeI &5270.356 &148.900 &3.780 &4.794\\
 FeI &5283.621 &49.100 &2.171 &4.375\\
\end{tabular}
\end{center}
\end{table}
 
 \begin{table}
\begin{center}
%\caption{Linelist for \sdsszero} \label{tbl-1}
\begin{tabular}{ccccccccccccrrrrrr}
\tableline
Species &$\lambda$ &obs.eqw &error\tablenotemark{*} &Abundance \\
 \tableline
 FeI &5307.361 &34.300 &1.814 &4.479\\
 FeI &5324.179 &69.500 &2.583 &4.403\\
 FeI &5332.900 &41.600 &1.998 &4.385\\
 FeI &5339.929 &28.000 &1.639 &4.194\\
 FeI &5393.168 &34.900 &1.830 &4.432\\
 FeI &5397.128 &140.200 &3.668 &4.309\\
 FeI &5405.775 &147.000 &3.756 &4.414\\
 FeI &5415.199 &29.888 &1.694 &4.348\\
 FeI &5434.524 &132.100 &3.561 &4.446\\
 FeI &5497.516 &89.501 &2.931 &4.384\\
 FeI &5501.465 &83.700 &2.834 &4.436\\
 FeI &5506.779 &83.400 &2.829 &4.252\\
 FeI &5576.089 &21.800 &1.446 &4.572\\
 FeI &5572.842 &44.000 &2.055 &4.274\\
 FeI &5624.542 &21.894 &1.450 &4.304\\
 FeI &6065.482 &39.500 &1.947 &4.410\\
 FeI &6136.615 &56.671 &2.332 &4.304\\
  FeI &6137.692 &50.000 &2.191 &4.383\\
 FeI &6230.723 &69.821 &2.589 &4.537\\
 FeI &6252.555 &58.000 &2.359 &4.572\\
 FeI &6265.134 &25.900 &1.577 &4.527\\
 FeI &6335.331 &28.700 &1.660 &4.317\\
 FeI &6358.698 &31.300 &1.733 &4.763\\
 FeI &6393.601 &53.451 &2.265 &4.230\\
 FeI &6421.351 &36.200 &1.864 &4.442\\
 FeI &6430.846 &62.100 &2.441 &4.605\\
\end{tabular}
\end{center}
\end{table}

\begin{table}
\begin{center}
%\caption{Linelist for \sdsszero} \label{tbl-1}
\begin{tabular}{ccccccccccccrrrrrr}
\tableline
Species &$\lambda$ &obs.eqw &error\tablenotemark{*} &Abundance \\
 \tableline
 FeI &6494.980 &78.500 &2.745 &4.429\\
 FeI &6592.914 &40.400 &1.969 &4.434\\
FeII &4233.172 &94.400 &3.010 &5.001\\
FeII &4491.405 &17.600 &1.300 &4.136\\
FeII &4583.837 &68.400 &2.562 &4.458\\
FeII &4923.921 &106.100 &3.191 &4.993\\
FeII &5018.440 &103.600 &3.153 &4.776\\
FeII &5197.568 &33.800 &1.801 &13.002\\
FeII &5316.615 &77.600 &2.729 &4.910\\
 NiI &4855.406 &8.100 &0.882 &3.040\\
 NiI &6643.629 &11.300 &1.041 &2.946\\
 NiI &6767.768 &16.200 &1.247 &3.165\\
 ZnI &4722.153 &11.500 &1.051 &1.668\\
 ZnI &4810.528 &9.300 &0.945 &1.434\\
\tableline
\end{tabular}
\end{center}
\begin{tablenotemark}
    \newline
    \tablenotemark{*} The errors are computed using Cayrel's relation \citep{cayrel1988}.
    \end{tablenotemark}
\end{table}

\begin{table}
\begin{center}
\caption{Linelist for \sdssone} \label{tbl-1}
\begin{tabular}{ccccccccccccrrrrrr}
\tableline\tableline
Species &$\lambda$ &obs.eqw &error\tablenotemark{*} &Abundance \\
 \tableline
  SiI &5268.387 &4.600 &0.664 &5.457\\
  SiI &6237.319 &4.200 &0.635 &5.358\\
  CaI &4226.728 & 112.700 &3.289 &3.221\\
  CaI &4283.011 &10.300 &0.994 &3.401\\
  CaI &4302.528 &23.800 &1.511 &3.368\\
  CaI &4318.652 &18.700 &1.340 &3.710\\
  CaI &4425.437 &27.400 &1.622 &3.910\\
  CaI &4454.779 &39.600 &1.950 &3.558\\
  CaI &4585.865 &12.200 &1.082 &3.679\\
  CaI &5857.451 &9.100 &0.935 &3.834\\
  CaI &6122.217 &28.700 &1.660 &3.959\\
  CaI &6162.173 &32.700 &1.772 &3.841\\
  CaI &6439.075 &21.600 &1.440 &3.677\\
  TiI &4533.241 &15.600 &1.224 &2.419\\
  TiI &4981.731 &13.100 &1.121 &2.288\\
  TiI &4991.065 &6.000 &0.759 &2.036\\
   TiI &4999.503 &9.800 &0.970 &2.368\\
  TiII &4012.385 &17.500 &1.296 &1.789\\
  TiII &4028.343 &21.100 &1.423 &2.522\\
  TiII &4163.648 &3.200 &0.554 &1.598\\
  TiII &4171.910 &3.500 &0.580 &1.739\\
  TiII &4290.219 &49.800 &2.186 &2.259\\
  TiII &4300.049 &53.300 &2.262 &1.899\\
  TiII &4312.864 &22.400 &1.466 &1.933\\
  TiII &4395.033 &58.600 &2.372 &1.888\\
  TiII &4443.794 &52.200 &2.238 &1.965\\
\end{tabular}
\end{center}
\end{table}
  
\begin{table}
\begin{center}
%\caption{Linelist for \sdssone}
\begin{tabular}{ccccccccccccrrrrrr}
\tableline\tableline
Species &$\lambda$ &obs.eqw &error\tablenotemark{*} &Abundance \\
 \tableline
  TiII &4468.507 &57.900 &2.357 &2.013\\
  TiII &4501.273 &48.400 &2.155 &1.974\\
  TiII &4533.969 &49.800 &2.186 &1.904\\
  TiII &4563.761 &36.600 &1.874 &1.887\\
  TiII &4571.968 &31.800 &1.747 &1.599\\
  CrI &4254.332 &54.600 &2.289 &2.187\\
  CrI &4274.796 &59.800 &2.396 &2.377\\
  CrI &4289.716 &46.200 &2.106 &2.240\\
  CrI &5204.506 &14.600 &1.184 &2.296\\
  CrI &5206.038 &31.900 &1.750 &2.504\\
  CrI &5208.419 &39.100 &1.937 &2.509\\
  CrI &4824.127 &4.300 &0.642 &2.834\\
  MnI &3823.507 &37.100 &1.887 &3.290\\
  MnI &4030.753 &41.100 &1.986 &1.622\\
  MnI &4033.062 &15.200 &1.208 &1.178\\
  MnI &4034.483 &17.900 &1.311 &1.449\\
  MnI &4823.524 &3.800 &0.604 &2.165\\
  FeI &4005.242 &72.350 &2.635 &4.325\\
  FeI &4045.812 & 109.200 &3.237 &4.339\\
  FeI &4132.058 &73.770 &2.661 &4.379\\
  FeI &4143.868 &77.740 &2.732 &4.299\\
  FeI &4187.039 &41.910 &2.006 &4.452\\
  FeI &4187.795 &36.600 &1.874 &4.310\\
  FeI &4198.304 &29.080 &1.671 &4.306\\
  FeI &4202.029 &73.499 &2.656 &4.140\\
  FeI &4227.427 &19.970 &1.384 &3.934\\
  FeI &4250.119 &31.760 &1.746 &4.095\\
\end{tabular}
\end{center}
\end{table}
  
  \begin{table}
\begin{center}
%\caption{Linelist for \sdssone}
\begin{tabular}{ccccccccccccrrrrrr}
\tableline\tableline
Species &$\lambda$ &obs.eqw &error\tablenotemark{*} &Abundance \\
 \tableline
  FeI &4250.787 &66.470 &2.526 &4.165\\
  FeI &4260.474 &63.579 &2.470 &4.213\\
  FeI &4271.154 &47.860 &2.143 &4.437\\
  FeI &4271.760 &92.450 &2.979 &3.977\\
  FeI &4325.762 &83.979 &2.839 &3.897\\
  FeI &4375.930 &41.550 &1.997 &4.511\\
  FeI &4383.545 & 112.400 &3.284 &4.508\\
  FeI &4404.750 &86.770 &2.886 &4.355\\
  FeI &4415.122 &69.280 &2.579 &4.522\\
  FeI &4427.310 &31.930 &1.751 &4.399\\
  FeI &4528.614 &38.710 &1.927 &4.440\\
  FeI &4531.148 &21.030 &1.421 &4.644\\
  FeI &4583.721 &25.720 &1.571 &7.103\\
  FeI &4871.318 &28.190 &1.645 &4.433\\
  FeI &4872.138 &26.150 &1.584 &4.603\\
  FeI &4891.492 &40.500 &1.972 &4.401\\
  FeI &4918.994 &26.150 &1.584 &4.360\\
  FeI &5006.119 &16.280 &1.250 &4.348\\
  FeI &5041.756 &13.310 &1.130 &4.420\\
  FeI &5171.596 &26.600 &1.598 &4.370\\
  FeI &5194.942 &20.500 &1.403 &4.603\\
  FeI &5216.274 &11.600 &1.055 &4.429\\
  FeI &5232.940 &38.020 &1.910 &4.387\\
  FeI &5266.555 &19.990 &1.385 &4.388\\
  FeI &5269.537 &76.789 &2.715 &4.240\\
  FeI &5324.179 &27.990 &1.639 &4.632\\
  FeI &5328.039 &79.800 &2.767 &4.483\\
  FeI &5328.532 &22.770 &1.478 &4.404\\
\end{tabular}
\end{center}
\end{table}

  \begin{table}
\begin{center}
%\caption{Linelist for \sdssone}
\begin{tabular}{ccccccccccccrrrrrr}
\tableline\tableline
Species &$\lambda$ &obs.eqw &error\tablenotemark{*} &Abundance \\
 \tableline
  FeI &5371.490 &67.410 &2.544 &4.445\\
  FeI &5397.128 &55.110 &2.300 &4.490\\ 
  FeI &5405.775 &58.820 &2.376 &4.487\\
  FeI &5429.697 &57.280 &2.345 &4.474\\
  FeI &5434.524 &36.540 &1.873 &4.386\\
  FeI &5446.917 &56.490 &2.328 &4.522\\
  FeI &5455.609 &45.910 &2.099 &4.532\\
  FeI &5572.842 &12.470 &1.094 &4.478\\
  FeI &5615.644 &24.400 &1.530 &4.603\\
  FeI &6230.723 &16.000 &1.239 &4.670\\
  FeI &6494.980 &18.970 &1.349 &4.552\\
  FeII &4233.172 &30.390 &1.708 &4.243\\
  FeII &4508.288 &15.630 &1.225 &4.494\\
  FeII &4515.339 &5.290 &0.713 &4.132\\
  FeII &4522.634 &7.240 &0.834 &3.890\\
  FeII &4555.893 &2.260 &0.466 &3.527\\
  FeII &4583.837 &25.720 &1.571 &4.294\\
  FeII &4923.927 &48.280 &2.153 &4.255\\
  CoI &4092.384 &7.900 &0.871 &2.267\\
  CoI &4121.311 &11.400 &1.046 &1.823\\
  NiI &3807.138 &30.300 &1.705 &2.603\\
  NiI &4401.538 &7.000 &0.820 &3.374\\
  NiI &4459.027 &22.000 &1.453 &4.258\\
  NiI &5476.900 &20.600 &1.406 &3.348\\

\tableline
\end{tabular}
\end{center}
\begin{tablenotemark}
    \newline
    \tablenotemark{*} The errors are computed using Cayrel's relation \citep{cayrel1988} .
    \end{tablenotemark}
\end{table}

%\bibliographystyle{apj}
%\bibliography{refer1.bib}

\end{document}